# Structural Dependency Analysis for Masked NTT Hardware: Scalable Pre-Silicon Verification of Post-Quantum Cryptographic Accelerators

Ray Iskander[1], Khaled Kirah[2,*]

[1]Verdict Security
[2] Faculty of Engineering, Ain Shams University, Cairo, Egypt

**Abstract**

Post-quantum cryptographic (PQC) accelerators implementing ML-KEM (FIPS 203) and ML-DSA (FIPS 204) require side-channel resistance evidence for FIPS 140-3 certification. However, exact masking-verification tools scale only to gadgets of a few thousand cells. We present a four-stage verification hierarchy, D0/D1 structural dependency analysis, fresh-mask refinement, Boolean Single-Authentication Distance Checking (SADC), and arithmetic SADC, that extends sound first-order masking verification to production arithmetic modules. Applied to the 1.17-million-cell Adams Bridge ML-DSA/ML-KEM accelerator, structural analysis completes in seconds across all 30 masked submodules. A multi-cycle extension (MC-D1) reclassifies 12 modules from structurally clean to structurally flagged. On the 5,543-cell ML-KEM Barrett reduction module, the pipeline machine-verifies 198 of 363 structurally flagged wires (54.5%) as first-order secure, reports 165 as candidate insecure for designer triage (a sound upper bound), and leaves 0 indeterminate. Every verdict is cross validated by Z3 and CVC5 with 0 disagreements across 363 wires. The result narrows manual review from hundreds of structural flags to 165 actionable candidates with mathematical certificates, enabling pre-silicon side-channel evidence generation on production ML-KEM hardware.

## 1. Introduction

The finalization of ML-KEM (FIPS 203) [1] and ML-DSA (FIPS 204) [2] in 2024 has intensified efforts to build side-channel-resistant hardware accelerators for lattice-based post-quantum cryptography. Masking remains the dominant countermeasure, decomposing each secret into multiple randomized shares. Security crucially depends on one invariant: no combinational logic path should allow two or more shares of the same secret to converge, because such a recombination creates an observable function of the secret under power consumption.

Verifying this invariant on a scale is difficult. Manual review does not scale. Trace-based Differential Power Analysis (DPA) testing requires fabricated hardware and can miss leaks that depend on specific data or rare control paths. Gate-level formal tools, SILVER [3], maskVerif [4], Prover [5], and Coco-Alma [6], provide rigorous single-cycle verification under the probing model [7] and have been highly successful for Boolean-masked symmetric primitives. Arithmetic masking over large fields, as required for Number Theoretic Transform (NTT)-based Post Quantum Cryptography (PQC), combined with deep pipelines creates a new challenge. Even when shares are separated within one clock cycle, pipeline registers and forwarding logic can create multi-cycle paths that recombine shares only after several stages. These longer dependency chains are invisible to any single-cycle analysis.

*Correspondence Author: khaled.kirah@eng.asu.edu.eg
Ray Iskander: ray@verdictsecurity.com

This paper introduces QANARY (**Q**uantum C**anary**), an early-warning tool that catches quantum-vulnerable cryptographic implementations before they reach production. Our tool takes its name from Canary birds which served for nearly a century as an early warning system in coal mines [8]. The tool is intended to be a structural dependency verification framework designed for large-scale arithmetic-masked NTT hardware. The method enforces a sound necessary condition for first-order probing security: structural non-convergence of shares. A wire is considered structurally insecure if its combinational fan-in depends on multiple shares of the same secret. Under the standard probing model, violation of this condition guarantees insecurity (zero false negatives). The converse does not hold, so algebraic independences produce unavoidable false positives. We position structural analysis as a scalable, sound-but-incomplete first stage that we then refine, in this paper, by two distributional checks (Boolean and arithmetic) that machine-verify a substantial fraction of the structurally-flagged wires as first-order secure. We make four contributions:

1. **Production-scale structural dependency verification.** We extend D0/D1 share-dependency analysis to the entire 1.17-million-cell Adams Bridge ML-DSA/ML-KEM accelerator (CHIPS Alliance / Caliptra project, 30 masked submodules), completing single-cycle analysis in 8.7 seconds. This establishes that structural verification, not only exact verification, must scale before any downstream refinement is possible at production sizes.
2. **Multi-cycle structural reclassification (MC-D1).** Conventional single-cycle dependency analysis stops at register outputs. MC-D1 extends this via fixed-point label propagation across cycle boundaries, reclassifying 12 of the 30 Adams Bridge modules from structurally clean (single-cycle) to structurally flagged (multi-cycle). These modules are surfaced as candidates for downstream verification, not as direct security verdicts. The Single-Authentication Distance Checking (SADC) refinement of contribution 3 has so far been applied to two of them: Domain-Oriented Masking (DOM) AND and Barrett reduction. For the NTT butterfly modules specifically, we report a definition-dependent disagreement between SADC (stable-state, single-probe) and Coco-Alma's per-location SAT mode in Section 4.6 rather than a unified security verdict.
3. **Four-stage hierarchy with arithmetic SADC.** On the 5,543-cell ML-KEM Barrett reduction module of Adams Bridge, our four-stage hierarchy machine-verifies 198 of 363 D0/D1-flagged wires (54.5%) as first-order secure under arithmetic value-independence, isolates 165 candidate first-order insecure wires as a sound upper bound for designer triage, and leaves 0 indeterminate, a complete circuit-level classification in approximately three minutes on a single core. Every arithmetic SADC verdict is independently confirmed by Z3 and CVC5 (363 of 363 wires, 0 disagreements). Methodologically, the hierarchy refines Boolean SADC verdicts via a value-independence test over a symbolic $S_0 = (X - S_1) \bmod q$ reparametrization, formalized as Theorem 3.9.1 (Section 3.9.4). To our knowledge, this is the first integration of an arithmetic-masking value-independence check into a Yosys-driven structural verification pipeline at production scale. Value-independence is sufficient but not necessary for distributional security, so the 165 wires are a strict over-approximation that an exact distributional check may further reduce.

4. **Dual-solver validation and reproducibility infrastructure.** Every arithmetic SADC query on Barrett is cross-checked by two independent SMT decision procedures: Z3 and CVC5 agree on all 363 of 363 wires with 0 disagreements, ruling out solver-implementation bugs as a source of any verdict. The pipeline is deterministic under pinned random seeds and resource-limit bounds, producing bit-identical headline numbers across runs. A pure- Python Z3-free Boolean evaluator is bundled for external ground- truth checking on small reference modules. Seventeen mandatory self-checks validate the SMT encoding of each analyzed module against reference wire functions, providing defense-in-depth against modeling mistakes. Limitations, including structural incompleteness and first-order-only scope, are detailed in Section 6.

Adams Bridge is extensively studied in literature. Independent vulnerability discoveries using power-trace DPA [9] and source-level inspection [10], identified overlapping share-combination patterns. The vendor's perspective on certain locations as intentional tradeoffs is presented in [11] and discussed in Section 5. Our contribution lies in the automated, RTL-to-netlist structural methodology that achieves exhaustive coverage pre-silicon. It complements rather than competes with trace-based and algebraic approaches.
This manuscript is organized as follows. Section 2 surveys related masking verification work. Section 3 presents the dependency model and MC-D1 algorithm. Section 4 reports evaluation results on Adams Bridge and standard benchmarks. Section 5 addresses vendor perspectives and complementarity of methods. Section 6 discusses limitations and research directions, and Section 7 concludes.

## 2. Literature Review

The probing model [7] provides the theoretical foundation for masking security. A circuit is d-probing secure if an adversary observing any d wires learns nothing about the secret. Verifying this property on real hardware has motivated a family of automated tools operating on gate-level netlists.

### 2.1. Formal Masking Verification Tools

We start by surveying available masking verification tools which operate primarily in single-cycle mode, verifying combinational logic within one clock cycle. Their evaluation targets are Advanced Encryption Standard (AES) S-boxes, Keccak permutations, and standard masking gadgets (DOM, ISW, HPC), typically at scales under 5,000 gates.
MaskVerif [4] was among the first automated tools for verifying masking security, checking Non-Interference (NI) and Strong Non-Interference (SNI) properties via symbolic computation [12]. It supports higher-order masking and the glitch-extended probing model, providing exact results within its specified model. REBECCA [13] introduced direct netlist verification in the glitch-extended probing model, formally accounting for transient values during gate evaluation. It supports limited sequential unrolling but targets primarily combinational analysis. SILVER [3] performs exact statistical independence verification using Reduced Ordered Binary Decision Diagrams (ROBDDs), providing results with no false positives at the cost of scalability. It verifies standard, glitch-extended, and Probe-Isolating Non-Interference (PINI) composability properties. Prover [5] extends SILVER's ROBDD approach with variable reduction rules and heuristic simulation-set enumeration, achieving faster verification while maintaining exact results [3]. It can verify S-boxes where SILVER times out. ProverNG [14] further extends this line with efficient compositional NI/SNI checking. IronMask [15] supports a wide range of security notions including random probing and composability, operating on abstract gadget descriptions rather than gate-level netlists. VerMI [16] verifies threshold implementation properties (non-completeness, uniformity) on hardware designs with inherent glitch awareness. INDIANA [17] provides exact verification through simulation-based indistinguishability analysis, scaling to full AES rounds.

We turn now to sequential and design-level tools. Coco-Alma [18] uses SAT-based analysis to verify masking security of sequential circuits, accounting for both glitches and transition leakage across clock cycles. Its execution-aware model captures cross-register leakage that single-cycle tools miss. Coco [19] extends this approach to full processor cores executing masked software, demonstrating verification of masked AES on the RISC-V Ibex core, the largest exact sequential verification for processor-scale designs to date. PROLEAD [20] takes a simulation-based approach, using statistical hypothesis testing on gate-level netlists to detect leakage. It scales to full cipher implementations (AES, Keccak, ASCON) and handles combined glitch-plus-transition leakage but provides statistical rather than formal guarantees.

aLEAKator [21] performs mixed-domain HDL simulation combining symbolic expression propagation with glitch and stability tracking. It verifies full masked AES implementations running on several CPU cores (Ibex, CV32E40P, Cortex-M3/M4) in 19–39 minutes, a significant advance in design-level verification. It requires a concrete execution stimulus and has been applied to Boolean masking for symmetric primitives. MATCHI [22] enables compositional verification of Hardware Private Circuit (HPC) designs, checking PINI and Output Probe-Isolating Non-Interference (OPINI) properties at composition boundaries. The approach requires HPC-annotated designs and cannot verify circuits that do not follow HPC discipline (e.g., DOM-based designs). MATCHI supersedes the earlier fullVerif [23] tool.

All gate-level tools above were developed for and validated on Boolean masking, where shares are related by XOR. Several handle sequential multi-cycle analysis (Coco-Alma, Coco, PROLEAD, aLEAKator, and MATCHI), though none has been applied to arithmetic masking in large-scale NTT-based post-quantum hardware. Additional tools targeting masking at the algorithm or code level include tightPROVE [24] for tight compositional probing security, VRAPS [25] for random probing security verification, CheckMasks [26] for formal gadget verification, and Tornado [27] for end-to-end masked implementation synthesis with built-in verification.

## 2.2. Arithmetic Masking Verification

While the tools in Section 2.1 excel in Boolean domains typical of symmetric cryptography, post-quantum schemes demand verification of arithmetic masking, where shares satisfy $s = (s_0 + s_1) \bmod q$ rather than $s = s_0 \oplus s_1$. Converting between arithmetic and Boolean domains (A2B/B2A) is well studied [28] [29].

Formal masking verification was extended to arithmetic operations, building on the Coco framework [30]. A2B and B2A conversion algorithms were verified with glitch awareness and supports up to second-order verification. While applicable to modular arithmetic relevant to PQC, it has not been demonstrated on full NTT pipelines or synthesized hardware netlists.

eVer [31] introduces a proof system for verifying masking security over arbitrary finite fields, extending to polynomial, inner-product, and code-based masking schemes. It has demonstrated operations over GF(3329) relevant to ML-KEM, though it operates at the algorithm level and has not been applied to synthesized hardware netlists. VERICA [32] supports combined leakage and fault-injection verification with an experimental arithmetic mode that remains untested in publications. A hand-proven PINI-secure hardware gadgets for prime-field arithmetic was presented [33], providing secure building blocks for masked PQC implementations. However, their security proofs are manual, not automated. IronMask has been extended to prime-field arithmetic (IronMaskArithmetic), but operates on abstract gadget descriptions, not gate-level netlists [15].
While Coco and eVer advance arithmetic masking verification, their scope does not extend to gate-level analysis of production NTT pipelines. No published tool has demonstrated automated gate-level verification of arithmetic masking in large-scale synthesized NTT hardware, where leakage mechanisms involve carry propagation in modular reduction, share mixing in butterfly data paths, and pipeline interactions across sequential stages in production designs.

### 2.3. Positioning of This Work

Table 1 compares our tool against existing masking verification approaches along the dimensions most relevant to our contributions: precision class, masking domain, maximum demonstrated scale, sequential (multi-cycle) support, and application to PQC hardware.

**Table 1. Comparison of masking verification tools.**

| Tool | Venue | Precision | Masking | Max Demonstrated Scale | Multi-cycle | PQC HW |
|---|---|---|---|---|---|---|
| maskVerif | ESORICS '19 | Exact (symbolic) | Bool | S-boxes ($d \leq 2+$) | No | No |
| REBECCA | EUROCRYPT '18 | Exact (netlist) | Bool | Gadgets (limited seq.) | Limited | No |
| SILVER | ASIACRYPT '20 | Exact (ROBDD) | Bool | S-boxes ($d \leq 3$) | No | No |
| Prover | TCHES '25 | Exact (ROBDD) | Bool | S-boxes ($d \leq 3$) | No | No |
| ProverNG | ICICS '25 | Exact (comp. ROBDD) | Bool | S-boxes ($d \leq 3$) | No | No |
| IronMask | S&P '22 | Exact (enumeration) | Bool | Gadgets ($d \leq 6$) | No | No |
| INDIANA | EUROCRYPT '25 | Exact (simulation) | Bool | Full AES rounds | No | No |
| Coco-Alma | FMCAD '21 | Exact (SAT) | Bool | ~2.3K cells | Yes | Partial[1] |
| Coco | USENIX Sec '21 | Exact (SAT/spectral) | Bool | Full Ibex CPU | Yes | No |
| PROLEAD | TCHES '22 | Statistical | Bool | AES, Keccak (100K+ gates) | Yes | No |
| aLEAKator | ePrint '25 | Exact | Bool | Masked AES on | Yes | No |

| Tool | Venue | Precision | Masking | Max Demonstrated Scale | Multi-cycle | PQC HW |
|---|---|---|---|---|---|---|
| | | (symbolic sim) | | CPUs | | |
| MATCHI | ePrint '25 | Exact (compositional) | Bool | Full AES ($d \le 4$)[2] | Yes | No |
| Gigerl et al. | ACNS '23 | Exact (SAT) | Both | A2B/B2A ($d \le 2$) | Yes | Partial[3] |
| eVer | ePrint '26 | Exact (proof system) | Both | Gadgets ($d \le 5$)[4] | No | No |
| **This work** | — | **Structural (sound)** | **Arith.** | **1.17M cells (30 modules)** | **Yes** | **Yes** |

*[1] Coco-Alma applied to a 2,304-cell NTT butterfly in this work (Section 4). [2] Requires HPC annotation; incompatible with some masking schemes (e.g., DOM). [3] A2B/B2A focus; not demonstrated on NTT operations. [4] Algorithm-level; no gate-level clock model. Demonstrated on GF(3329).*

Our structural dependency verification trades completeness for scale and arithmetic/NTT coverage, providing a sound overapproximation: a wire flagged insecure may be algebraically independent of the secret (false positive), but no truly insecure wire is missed (zero false negatives under the standard probing model). Residual false positives from D0/D1 are addressed by a four-stage verification hierarchy (D0/D1 → FM → Boolean SADC → Arithmetic SADC, section 3.9), which eliminates 100% of structural false positives on the DOM AND reference gadget, and on the 5,543-cell ML-KEM Barrett reduction module machine-verifies 198 of 363 D0/D1-flagged wires (54.5%) as first-order secure under arithmetic value-independence and isolates 165 candidate-insecure wires as a sound upper bound, with 0 indeterminate. On arithmetic carry chains of the NTT butterfly extractions, SADC's stable-state single-probe verdict differs from a transient + glitch-loose Coco-Alma run, a definition-dependent observation we will investigate in Section 4.6. For modules within the reach of exact tools, typically under 5,000 cells, SILVER, Prover, or Coco-Alma provide strictly stronger guarantees. Our strength is production-scale multi-cycle analysis: MC-D1 detects cross-register share convergences through fixed-point label propagation on netlists totaling 1.17 million cells, a scale where exact multi-cycle tools like Coco-Alma and aLEAKator, while capable of sequential verification, have not been demonstrated on arithmetic-masked PQC designs. Section 3 details the methodology; Section 5 quantifies the precision trade-offs.

## 3. Methodology

We present a four-stage hierarchy for structural dependency analysis of masked hardware: single-cycle analysis (SC-D1), multi-cycle extension (MC-D1), fresh masking refinement (FM), and the Statistical Algebraic Distributional Check (SADC) for both Boolean and arithmetic masking. This Section defines each stage, argues soundness as a necessary condition for probing security, and describes the tool pipeline and self-validation infrastructure. Figure 1 shows the label lattice used throughout this analysis.

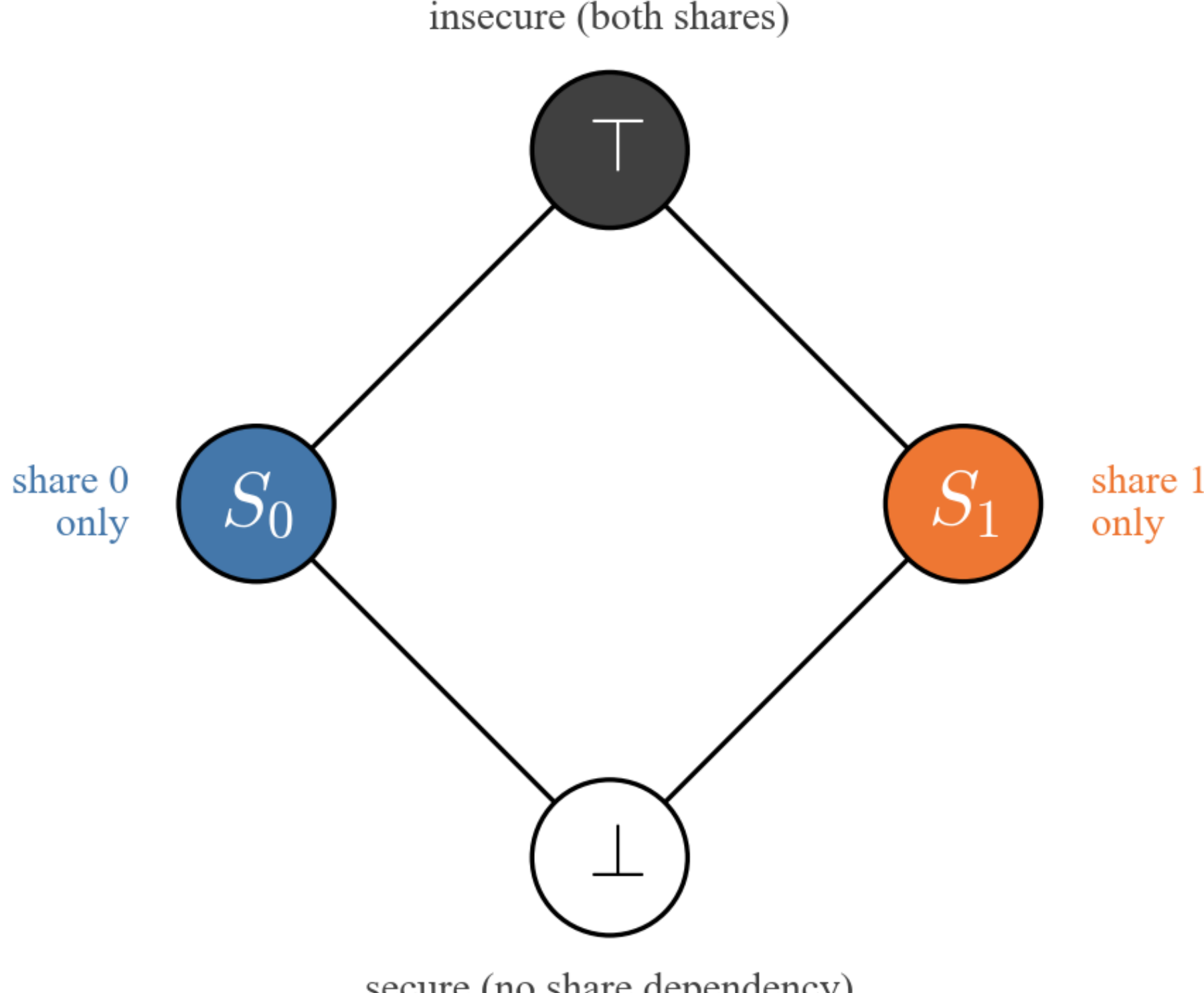

*Figure 1. Label lattice L = {⊥, $S_0$, $S_1$, BOTH} for first-order (d = 1) structural dependency analysis.*

### 3.1. Problem Statement

Consider a first-order masking scheme where a secret value s is split into two shares s_0 and s_1 such that s = s_0 ⊕ s_1 (Boolean masking) or s = (s_0 + s_1) mod q (arithmetic masking). Under the standard probing model [7], a circuit is first-order probing secure if an adversary observing any single wire learns nothing about the secret s.

We verify the necessary condition for the structural share separation. A wire *w* in the circuit has a combinational fan-in, the set of primary inputs that can influence *w* through combinational logic. If this fan-in includes inputs from at most one share group, the wire cannot jointly depend on both shares and is therefore independent of the secret. If the fan-in includes inputs from both share groups, the wire fails the necessary condition of share separation. Algebraic protections (fresh randomness, cancellations) may still ensure security, but structural analysis cannot confirm this.

Formally, let *w* = *f* (s_0, s_1, r, p) be the Boolean function computed at wire *w*, where r denotes fresh randomness and p denotes public inputs. For arithmetic masking, Z3 bit-blasts modular arithmetic (mod q) into Boolean logic; the queries below operate on the resulting Boolean functions. We define two dependency queries:

**D_0(w):** ∃ s_0, s_0’, s_1, r, p : s_0 ≠ s_0’ ∧ *f* (s_0, s_1, r, p) ≠ *f* (s_0’, s_1, r, p)

**D_1(w):** ∃ s_0, s_1, s_1’, r, p : s_1 ≠ s_1’ ∧ *f* (s_0, s_1, r, p) ≠ *f* (s_0, s_1’, r, p)

Each query asks: does there exist an input assignment where varying one share group alone changes the output? If SAT, the wire functionally depends on that share group. If UNSAT, the wire is functionally independent of that share group, equivalently, ∀ s_i, s_i' : *f* (..., s_i, ...) = *f* (..., s_i', ...) for all other inputs fixed. For multi-bit share groups, each query is evaluated over the full bit-vector; Z3 bit-blasts the vector comparison into per-bit constraints.

A wire is classified SECURE if D_0 or D_1 returns UNSAT (depends on at most one share group), and POTENTIALLY_INSECURE if both return SAT (depends on both share groups). The tool reports UNKNOWN if the SMT solver times out on either query.

### 3.2. Soundness Argument

Structural share separation is a sound necessary condition for first-order probing security:

Theorem (Soundness). Under the standard probing model, assuming glitch-free combinational evaluation, uniform and mutually independent shares and randomness, and single-wire observation, if a wire *w* is structurally independent of at least one share group (D_0 or D_1 is UNSAT), then observing *w* reveals no information about the secret s.
*Proof sketch.* If D_0 is UNSAT, then w = g(s_1, r, p) for some function g, the wire's output is determined entirely by s_1, r, and p. Because s_0 is chosen uniformly at random and kept secret, any fixed value of s induces a uniform distribution over s_1 (Boolean case: s_1 = s ⊕ s_0) or over the residue class (s - s_0) mod q (arithmetic case). Thus w = g(s_1, r, p) is statistically independent of s, i.e., P(w | s) = P(w) for all s, and therefore I(s; w) = 0. The argument for D_1 is symmetric.
This soundness holds under the stated assumptions. Extensions to glitch-extended probing [34], transition leakage [35], or higher-order (multi-wire) attacks require modeling physical phenomena beyond structural dependency. The analysis does not model physical effects such as glitches, coupling, or transition leakage. These are discussed in Section 6.
The converse does not hold. A wire may depend on both s_0 and s_1 structurally yet be statistically independent of s due to algebraic cancellations (e.g., re-masking with fresh randomness). This is the source of false positives. On a reference DOM AND gadget [36], structural analysis flags wires as insecure that are provably statistically independent of the secret. The four-stage hierarchy of section 3.9 closes this gap exactly on DOM AND, after the SADC stage all flagged wires are either machine-verified secure or confirmed insecure; the residual false-positive rate at the end of the pipeline is 0%.

### 3.3. Tool Pipeline

The verification pipeline processes hardware designs in five stages:

1. **RTL Conversion.** SystemVerilog source is converted to Verilog using sv2v [37]. This step resolves SystemVerilog constructs (interfaces, structs, parameterized modules) into synthesizable Verilog.
2. **Synthesis.** Yosys [38] synthesizes the Verilog to a gate-level netlist: synth -flatten; techmap; dffunmap; opt. The dffunmap pass decomposes complex flip-flop cells into primitive D-type flip-flops ($_DFF_P_, $_DFF_N_, etc.) that the adapter can enumerate. The resulting netlist contains only combinational gates ($_AND_, $_OR_, $_XOR_, $_NOT_, $_MUX_) and primitive DFFs.
3. **Netlist Adaptation.** The NetlistAdapter parses the Yosys JSON netlist and builds per-wire Z3 expressions by topological propagation. Module hierarchy is fully flattened during synthesis (stage 2) so that inter-module combinational paths are correctly captured. DFF outputs (Q pins) are treated as free variables, combinational cut points that decouple sequential stages. Each wire's Boolean function is expressed in terms of primary inputs (s_0, s_1, r, p) and DFF free variables.

4. **Dependency Analysis.** The ProbingVerifier evaluates D_0 and D_1 queries for each wire using Z3 [39] with a per-query timeout of 10 seconds. A JSON configuration file maps port names to share groups (s_0, s_1, r, p), defining the input partition for each target module.
5. **Report Generation.** Results are aggregated into a CircuitReport: per-wire verdicts, per-module statistics, and overall classifications (CLEAN, INSECURE, or TIMEOUT).

### 3.4. Complexity Analysis

For a netlist with |V| wires and |E| gate-input edges, single-cycle dependency propagation requires O(|V| + |E|) set-union operations on the dependency lattice, one topological pass. MC-D1 adds a fixed-point loop bounded by D iterations (longest acyclic DFF chain depth), giving O(D · (|V| + |E|)) total work. Since the lattice has height 3, each wire's label changes at most 3 times across all iterations, so the effective cost is O(|V| + D · |E_DFF|) where |E_DFF| is the number of DFF boundary edges. Each D_0/D_1 SMT query operates on a wire's Boolean function, whose size is bound by the combinational cone depth. Z3 resolves individual queries in microseconds for cones under 10,000 gates; larger cones may require milliseconds due to bit-blasting complexity.

For the Adams Bridge accelerator, SC-D1 analyzes 27 modules (1.17 million cells) in 8.7 seconds total. MC-D1 completes in 231.8 seconds across all 27 modules (DFF chain depths 0–263). Runtime varies by module: small modules (under 1,000 cells) complete in under 0.1 seconds, while the largest completed module (238,000 cells) requires 2.1 seconds for SC-D1 and 128.7 seconds for MC-D1 (263 iterations).

### 3.5. Self-Checks

Before analyzing any target circuit, the tool executes 17 mandatory known-answer checks that validate the SMT encoding against circuits with known security properties. These were introduced after a tautological encoding failure during early development (an over constrained solver returned UNSAT for all queries, producing false SECURE verdicts). The checks cover:

- 7 masked Boolean circuits: isolated shares (SECURE), combined shares (INSECURE), XOR-masked products (INSECURE, structural FP by design), public-controlled MUX (INSECURE), and constant outputs (SECURE).
- 4 unmasked circuits: secret-dependent wires (INSECURE) and public-only wires (SECURE).
- 6 arithmetic-mode circuits: modular reduction outputs, carry propagation chains, and Barrett-specific patterns.

All 17 checks must pass before any target analysis proceeds. A single failure halts execution. This is an engineering safeguard; no self-check has caught a runtime bug beyond the initial encoding error that motivated their creation.

### 3.6. Reproducibility

All empirical experiments in Section 4 use pinned tool versions: Python 3.10.12, Z3 4.13.0, Yosys 0.47, sv2v 0.0.12; the algebraic proof suite of Section 3.9.5 uses a more recent solver stack (Z3 4.15.4 + CVC5 1.3.3) documented in §4.5 Reproducibility, with both configurations pinned in `repro_manifest.json`. The Adams Bridge target is fixed at commit cdc9b1c (chipsalliance/adams-bridge). SHA-256 hashes for all 30 synthesized netlists are provided in the supplementary material.

The tool is distributed as a Python package with a command-line interface: qdebug formal verify <module.json> --label <labels.json>. A Docker image with all dependencies is available for deterministic reproduction. The test suite (1,729 tests, 79% line coverage) runs in under 4 minutes. We intend to submit to the TCHES artifact evaluation process.

**Code and Data Availability.** The complete reproduction artifact is publicly available at https://github.com/rayiskander2406/qanary-structural-dependency-analysis-arXiv-2604.15249 under the Apache-2.0 license and permanently archived on Zenodo (concept DOI https://doi.org/10.5281/zenodo.19625392; version 1.0.0 DOI https://doi.org/10.5281/zenodo.19625393). The repository includes the single-command reproduction driver (reproduce.py), the Z3/CVC5 proof suite (theorems T1–T6), per-module evidence JSON for all 27 synthesized modules, synthesized netlists pinned to Adams Bridge commit cdc9b1c, and repro_manifest.json with pinned tool versions and SHA-256 hashes.

### 3.7. Multi-Cycle Extension (MC-D1)

Single-cycle analysis treats DFF outputs as free variables, capturing only combinational dependencies within one clock cycle. This misses cross-register share convergence: if share s_0 is registered in cycle t and share s_1 in cycle t+1, their convergence at a downstream gate produces a structurally insecure wire invisible to single-cycle analysis. In the Adams Bridge NTT butterfly module, single-cycle analysis reports zero insecure wires; the vulnerability manifests only when shares propagate across register boundaries.
MC-D1 extends structural dependency analysis to multiple clock cycles via fixed-point label propagation. This technique is standard in compiler dataflow analysis [40] [41]. We apply Jacobi-style fixed-point iteration to masked hardware netlist verification, propagating share-dependency labels across DFF boundaries until convergence. The contribution is not the iteration technique itself, but its application to arithmetic masking verification at production scale, as demonstrated in Section 4 on the 1.17-million-cell Adams Bridge accelerator.

**Dependency lattice.** Each wire is assigned a label from the join-semilattice L = {⊥, S_0, S_1, BOTH}, ordered by ⊥ ⊑ S_0 ⊑ BOTH and ⊥ ⊑ S_1 ⊑ BOTH. The join operation is set union: a gate whose inputs carry labels S_0 and S_1 produces output label BOTH. The lattice has finite height 3.

**Algorithm.** MC-D1 proceeds in three phases:

*Phase 1: Initialization.* Primary inputs are labeled according to the input configuration (s_0 → S_0, s_1 → S_1, r → ⊥, p → ⊥). DFF Q-outputs are initialized to ⊥ (no dependency).

*Phase 2: Combinational propagation.* Gates are processed in topological order. Each gate's output label is the join of its input labels.

*Phase 3: Fixed-point iteration.* For each DFF, the label at its D-input (from Phase 2) is compared to its Q-output label. If they differ, the Q-output is updated. All DFF updates are applied atomically per iteration (Jacobi snapshot semantics), ensuring each iteration corresponds to exactly one clock-cycle boundary crossing. Phase 2 is then repeated. The algorithm terminates when no DFF label changes. Note that glitch effects are modeled only within a single combinational stage (via the join of all gate-input labels); registers terminate glitch propagation, but the labels resulting from within-cycle glitches propagate to subsequent cycles via the DFF D-input.
The maximum iteration count D is computed as the longest path in the DFF-level dependency graph after condensing strongly connected components (SCCs) via Tarjan's algorithm. DFFs within an SCC (feedback loops) are treated as cut points: they retain their accumulated labels, which is a sound overapproximation (labels can only grow, never shrink).

**Algorithm 1:** MC-D1 Fixed-Point Label Propagation

**Input**: Netlist **N**, input labels **L_in**
**Output**: Per-wire dependency labels

```
 1: D ← longest_path(condense_SCC(DFF_graph(N)))
 2: for each primary input i:
    3: label[i] ← L_in[i]
 4: for each DFF q-output:
    5: label[q] ← ⊥
 6: propagate_combinational(N, label)
 7: for k = 1, 2, …, D + 1:
    8: snapshot ← copy(label)
    9: changed ← false
   10: for each DFF (d_in, q_out) in N:
      11: if label[q_out] ⊏ snapshot[d_in]:
         12: label[q_out] ← snapshot[d_in]
         13: changed ← true
   14: if not changed: return label  ▷ Converged
   15: propagate_combinational(N, label)
16: return label
```

**Proposition (Bounded Convergence).** MC-D1 converges in at most D + 1 iterations, where D is the longest acyclic path in the SCC-condensed DFF-level dependency graph.

*Proof sketch.* Labels are drawn from a join-semilattice of height 3 and can only grow (monotonicity). Each iteration propagates labels through one additional DFF boundary. After D iterations, labels from the deepest pipeline input have reached all reachable DFFs. The (D+1)-th iteration detects convergence (no label changes). □

On the 27 Adams Bridge modules, MC-D1 converges in exactly D + 1 iterations for 25 modules, with 2 converging 1 iteration early due to parallel independent paths. DFF chain depths range from 0 to 263 (median 7). Total runtime: 231.8 seconds.

MC-D1 preserves the soundness guarantee: labels only grow under monotonic join, so if a wire's label never reaches BOTH, it is structurally separated from the secret across all pipeline depths. The overapproximation grows with pipeline depth, additional false positives arise as more share labels propagate through DFF chains. We quantify this trade-off in Section 5.

### 3.8. Fresh Masking Refinement

As an optional post-processing step, we test whether structurally insecure wires are re-masked by fresh randomness. For each wire w classified POTENTIALLY_INSECURE, and for each randomness bit r_i in w's fan-in, we check whether r_i acts as a bijection on w:

$\forall\ s_0, s_1, r_\{\neg i\}, p : w(r_i = 0) \neq w(r_i = 1)$

This is implemented by checking the negation for unsatisfiability. If UNSAT, flipping r_i always flips w regardless of other inputs. For single-bit wire outputs, this bijection property implies that w is uniformly distributed when r_i is uniform (assuming the restricted function is surjective, which holds for all standard linear and switching gates used in masking), yielding I(s; w) = 0 — r_i acts as a one-time pad.

On the Adams Bridge single-cycle analysis, FM promotes 1,316 of 64,038 structurally insecure wires to SECURE (2.05% reduction) at approximately 500× the cost of SC-D1. The 97.95% remaining indeterminate reflects DFF cut-point boundaries (invisible to combinational FM) and non-bijective gate outputs. FM addresses the subset of residual flags where a single fresh randomness bit acts as an XOR mask; it cannot resolve flags where security derives from the randomness of the shares themselves, the gap closed by the distributional checks of §3.9.

### 3.9. Statistical Algebraic Distributional Check

FM refinement leaves residual structural flags where the share bits are themselves uniform random and no fresh randomness bit appears in the wire's combinational fan-in. The canonical example is the cross- domain partial product $a_0 \wedge b_1$ in a DOM AND gadget: the wire structurally depends on a share of secret $a$ and a share of secret $b$, yet its output distribution is independent of both, because $a_0$ is a uniform mask for one secret and $b_1$ is an independent uniform mask for the other. FM cannot promote such wires; an exact distributional check is required. SADC provides this check as a pipeline stage that operates on the same Yosys-synthesised netlist as D0/D1 and FM.

#### 3.9.1. The reparametrization insight

D0/D1 treats the shares $(s_0, s_1)$ as adversarial: it varies them freely and asks whether the wire value can change. Under masking, however, shares are *not* adversarial; they are jointly distributed with the secret. For Boolean masking $s_0 = x \oplus s_1$ where $s_1$ is a uniform Boolean mask drawn independently of the secret $x$. For arithmetic masking modulo $q$, $s_0 = (x - s_1) \bmod q$ where $s_1$ is uniform on $\mathbb{Z}_q$. A wire observed by a first-order probe is first-order secure if and only if its output distribution (over all mask and fresh randomness, with the secret held fixed) is independent of the secret.

SADC reparametrizes paired share bits in terms of the underlying secret and an independent mask, then checks whether the wire's value *as a function of the secret* admits any secret-dependent assignment. This converts a structural fan-in question into an algebraic satisfiability question that the SMT solver can resolve directly.

### 3.9.2. Boolean SADC

Let $w(s_0, s_1, r)$ be the bit-vector function computed at a flagged wire under Boolean masking, where $r$ denotes fresh randomness. Let $\mathcal{C}_{s_0}, \mathcal{C}_{s_1}, \mathcal{C}_r$ denote the bit-indices of $s_0, s_1, r$ that appear in the wire's combinational fan-in. Boolean SADC introduces a fresh symbolic secret $x$ over the bits $\mathcal{C}_{s_0} \cap \mathcal{C}_{s_1}$ (the *paired* bits, where both shares of the same position appear in the cone) and substitutes $s_0[i] \mapsto x[i] \oplus s_1[i]$ for each paired index $i$. Unpaired share bits are treated as independent uniform randomness. The check then asks Z3 whether there exist two distinct secret assignments $x \neq x'$ that produce different wire values for some $(s_1, r)$:

$$\exists\, x, x', s_1, r \,:\, x \neq x' \;\wedge\; w_x(s_0 \leftarrow x \oplus s_1,\, s_1,\, r) \;\neq\; w_{x'}(s_0 \leftarrow x' \oplus s_1,\, s_1,\, r).$$

UNSAT yields a Boolean-SADC-secure verdict (the wire value is independent of the secret pointwise, and therefore distributionally). SAT yields a Boolean-SADC-insecure witness $(x, x', s_1, r)$ that produces the differing values. INDETERMINATE is reserved for solver resource-limit exhaustion (default per-wire query rlimit: $2 \times 10^7$).

### 3.9.3. Arithmetic SADC

Boolean SADC is exact for Boolean-masked wires but mismodels arithmetic-masked circuits, where the secret-share relation is modular subtraction rather than XOR. Under arithmetic masking the high bits of $s_0$ are coupled to the low bits of $s_1$ via the borrow chain, and a Boolean SADC run that touches high bits of either share cannot faithfully express the secret. On Adams Bridge Barrett under the deterministic configuration of §3.6, Boolean SADC resolves 146 of the 363 D0/D1-flagged wires as confirmed insecure (via low-bit leaks) and leaves 217 INDETERMINATE because the mask cone exceeds the Boolean enumeration budget.

Arithmetic SADC closes this residual by performing the reparametrization symbolically over the actual modular semantics. Let $X, X' \in \mathbb{Z}_q$ be candidate secrets, $S_1 \in \mathbb{Z}_q$ a uniform mask, and let

$$S_0 \;=\; (X - S_1) \bmod q, \qquad S_0' \;=\; (X' - S_1) \bmod q$$

be derived as full-width symbolic bitvectors of width $w$ matching the Adams Bridge netlist convention ($w = 24$ for both ML-KEM and ML-DSA Barrett, satisfying the no-overflow precondition $2q < 2^w$ that T4 verifies; ML-KEM has $2q = 6{,}658$ with comfortable headroom, while ML-DSA is tight at $2q = 16{,}760{,}834 < 2^{24} = 16{,}777{,}216$ , leaving 16{,}382 units of slack). Per-bit views $s_0[i] = \text{Extract}(i, S_0)$ and $s_1[i] = \text{Extract}(i, S_1)$ are then substituted into two copies of the wire expression to obtain $w_X$ and $w_{X'}$. The arithmetic value- independence check asks:

$$\neg\, \exists\, X, X', S_1, r \,:\, X < q \;\wedge\; X' < q \;\wedge\; S_1 < q \;\wedge\; X \neq X' \;\wedge\; w_X \neq w_{X'}.$$

UNSAT promotes the wire to arithmetic-SADC-secure. SAT yields a witness; the wire is reported as arithmetic-SADC-INSECURE_CONSERVATIVE (see §3.9.6 for why this is a sound *upper* bound rather than a definitive insecurity verdict). The runtime overflow assertion $2q < 2^w$ is checked once per pass and prevents the modular subtraction from wrapping the bitvector silently, on Adams Bridge ML-KEM Barrett, $q = 3{,}329$ and $w = 24$, well within the bound.

### 3.9.4. Soundness

**Definition 3 (value-independence under masking).** Let $w(s_0, s_1, r)$ be a bit-vector function and let $(s_0, s_1)$ jointly mask a secret $x$ via either Boolean reparametrization $s_0 = x \oplus s_1$ or arithmetic reparametrization $s_0 = (x - s_1) \bmod q$. Wire $w$ is value-independent of $x$ if, for every fixed assignment to the mask $s_1$ and the fresh randomness $r$, the value of $w$ does not change as $x$ varies over its domain.

**Theorem 3.9.1 (value-independence implies first-order distributional security).** Assume:

(A1) the adversary observes a single bit-vector wire (first-order single-probe);

(A2) the mask $s_1$ is drawn uniformly from its declared domain ($\{0,1\}^k$ for Boolean, $\mathbb{Z}_q$ for arithmetic) independently of the secret $x$;

(A3) the fresh randomness $r$ is drawn uniformly and is independent of both the secret and the mask;

(A4) all randomness channels labeled in the input port map are mutually independent.

Then if $w$ is value-independent of $x$ in the sense of Definition 3, the conditional distribution $\Pr[w \mid x]$ is constant in $x$ and hence $I(x; w) = 0$ — the probe leaks no information about the secret.

*Proof sketch.* Pointwise independence ($w(x, s_1, r) = w(x', s_1, r)$ for all $s_1, r$ and all $x, x'$) implies that the function $x \mapsto w$ is constant for any fixed $(s_1, r)$. The marginal distribution of $w$ at any fixed $x$ is therefore the pushforward of the uniform distribution on $(s_1, r)$ through this constant map, which by (A2)–(A4) does not depend on $x$, (A2) and (A3) supply the uniformity of the mask and the fresh randomness independently of $x$, while (A4) is what licenses factoring the joint distribution of $(s_1, r)$ into its marginal components above. Mutual information then vanishes by (A1): a single-probe adversary's observation is exactly $w$, so a constant conditional distribution implies $I(x; w) = 0$. ▫

The converse does not hold: a wire whose value *does* change with the secret for some $(s_1, r)$ may still admit a constant marginal distribution after averaging over $(s_1, r)$, threshold implementation rebalancing is the canonical example. Arithmetic SADC flags such wires as INSECURE_CONSERVATIVE; an exact distributional analysis (e.g., model counting) can promote some of them to secure.

### 3.9.5. Implementation and dual-solver validation

SADC runs as a refinement pass over the FM output. For each wire that enters as INDETERMINATE or insecure under D0/D1+FM, the implementation:

1. analyses the randomness cone via the same NetlistAdapter used by D0/D1, identifying $\mathcal{C}_{s_0}$, $\mathcal{C}_{s_1}$, and $\mathcal{C}_r$ from the per-wire Z3 expression;
2. partitions $\mathcal{C}_{s_0}$ into *paired* indices (the corresponding $s_1$ partner is also in the cone) and *unpaired* indices (treated as independent uniform randomness);
3. introduces fresh symbolic secret bitvectors $X, X'$, derives the reparametrized $S_0, S_0'$ (Boolean: XOR, arithmetic: modular subtraction), and substitutes them into two copies of the wire expression;
4. asserts $X \neq X'$ (and the range constraints $X, X', S_1 < q$ in the arithmetic case) and checks whether any value-distinguishing assignment exists.

The implementation uses Z3 (de Moura and Bjørner, TACAS 2008) as the primary solver with deterministic configuration: pinned global random seeds (`smt.random_seed=0`, `sat.random_seed=0`), per-query resource limits (`rlimit=2 \cdot 10^7` for Boolean, $10^7$ for arithmetic), and wall-clock timeouts retained only as a fallback. Under this configuration the headline numbers are bit-reproducible across runs.

Every arithmetic SADC query on the Adams Bridge ML-KEM Barrett module is additionally cross-checked by CVC5 [42]. The Z3 solver state is exported as SMT-LIB2 via `solver.to_smt2()` and re-parsed by CVC5's input parser; the second result is compared to the first. Across all 363 flagged Barrett wires, Z3 and CVC5 agree on every verdict with 0 disagreements. This rules out solver-implementation bugs as a source of any individual SADC outcome, the two solvers use independent decision procedures, and consistent agreement on a query of this size is strong evidence that the SMT encoding faithfully expresses the intended semantics.

**Formal proof suite for the Section 3.9 algebraic backbone.** Beyond the per-wire dual-solver validation reported above, the algebraic backbone of Section 3.9, a small-domain finite expansion of Theorem 3.9.1 over $q = 5$ together with the supporting lemmas T2 through T6, is machine-checked by Z3 and CVC5 in a multi-theory SMT proof suite distributed with the artifact (`scripts/proofs/`). T2 verifies the Boolean reparametrization round-trip $(x \oplus s_1) \oplus s_1 = x$ over 24-bit shares; T3 verifies the arithmetic reparametrization round-trip $\text{URem}(\text{URem}(x - s_1 + q, q) + s_1, q) = x$ for both deployed configurations: ML-KEM ($q = 3329$, $w = 24$) and ML-DSA ($q = 8\,380\,417$, $w = 24$, the tight 1-bit-slack case), matching the exact encoding used by `sadc_arith.py:352`. T4 verifies the no-overflow precondition $1 \leq (x - s_1 + q) < 2q < 2^w$ for all three configurations: ML-KEM ($w = 24$, comfortable headroom), ML-DSA ($w = 24$, tight 16{,}382-unit headroom, the deployed case), and ML-DSA ($w = 46 = 2\lceil \log_2 q \rceil$, included as a conservative-bound sanity check). T5 verifies via exhaustive Python enumeration over the entire 4096-element 12-bit RNG domain, with Z3 and CVC5 cross-checking the closed-form $\lceil (N - r)/q \rceil$ predicate for self-consistency, that the maximum bias ratio is 2 for raw 12-bit RNG reduced into $\mathbb{Z}_{3329}$, anchoring the mask-uniformity caveat in Section 6 L9. T6 enumerates a worked instance of Theorem 3.9.1 over $q = 5$ that machine-checks both directions of the soundness/tightness gap, a value-independent wire whose marginal is constant in $x$ (the theorem) and a non-value-independent wire whose marginal is *also* constant in $x$ (the canonical threshold-rebalancing example justifying the INSECURE_CONSERVATIVE label rather than INSECURE). T1 then machine-checks an *r-free sub-theorem* of Theorem 3.9.1, universally quantified over the entire $2^{25}$-element Boolean wire-function space at $q = 5$ for every secret pair, in which the fresh-randomness channel $r$ is collapsed; the lift from this r-free formalization to the r-bearing Theorem 3.9.1 follows from the law of total probability under (A3) and (A4) and is documented in T1's docstring, but is not itself verified by SMT. CVC5 acts as the universal prover for the T1 obligation and Z3 cross-checks. Adopting the multi-theory SMT structure of [42], the suite scopes its claims to the algebraic backbone of the methodology rather than to every empirical statement in the paper; the per-wire SADC verdicts on Adams Bridge are validated empirically by the 363/363 dual-solver agreement reported above. Both solvers agree on every proof obligation in the suite (0 disagreements across all six proofs), and the entire suite runs in a few seconds on a 2024 Apple Silicon laptop (the dominant cost is solver subprocess cold-start, not proof search; the verdicts are invariant across runs regardless of wall-clock variation). A designer skeptical of the arithmetic SADC encoding can therefore reproduce every algebraic claim of Section 3.9 in a few seconds without working through any proof sketches by hand, to our knowledge this is the first masking-verification artifact to ship a machine-checked proof suite for the algebraic foundations of its SMT encoding. SILVER [3] ships with Isabelle/HOL correctness proofs of its ROBDD verification framework, which target a different level of the stack: the soundness of binary decision diagrams as a representation for the probing-security check. Our proof suite targets the soundness of the SMT encoding used by arithmetic SADC, specifically, the no-overflow precondition, the arithmetic reparametrization round-trip, and a small-domain finite expansion of the value-independence theorem. The two efforts are complementary, not overlapping. The fully-universal version of Theorem 3.9.1 (quantifying simultaneously over all primes $q$ and the entire wire-function space) is reported as future work in Section 6.

### 3.9.6. Hierarchy and scope

SADC completes the four-stage hierarchy

$$\text{D0/D1} \subseteq \text{FM} \subseteq \text{Boolean SADC} \subseteq \text{Arithmetic SADC} \subseteq \text{Exact verification}$$

where each stage is at least as precise as the previous, and the arithmetic stage acts as a bridge between scalable structural screening and per-wire exact verification. All four stages operate inside a unified SMT-based flow over the same flattened Yosys netlist.

To our knowledge this is the first integration of an arithmetic-masking value-independence check into a Yosys-driven structural verification pipeline at production scale. Boolean distributional checks have appeared in earlier work, most notably SILVER's exact ROBDD analysis [3] and Prover's variable-reduction extension [5], but those tools operate as stand-alone exact verifiers on small gadgets and do not, in their published form, integrate as a refinement layer over a structural pre-screener operating on million-cell netlists. The novelty here is not the distributional check itself but its placement as the final stage of a screening hierarchy that scales to production arithmetic modules.

Arithmetic SADC's value-independence is sufficient but not necessary for distributional first-order probing security. A wire reported INSECURE_CONSERVATIVE may still be distributionally secure under exact model counting, threshold- implementation rebalancing, or compositional refresh masking. The INSECURE_CONSERVATIVE count is therefore a sound upper bound on the genuinely first-order insecure set; it never under-reports leakage but may over-report it. We treat the gap between value- independence and distributional independence as the principal direction for future SADC extensions (Section 6).

## 4. Evaluation

We evaluate our structural dependency analysis tool (v0.7.2, commit af97ec3) on the Adams Bridge masked PQC accelerator, the largest open-source NTT implementation with first-order masking. We report single-cycle results (SC-D1), multi-cycle results (MC-D1), standard benchmarks for generalizability, independent confirmation with Coco-Alma, and a comparison with existing masking verification tools. Full per-module results appear in Table 6.

### 4.1. Target: Adams Bridge

Adams Bridge (CHIPS Alliance / Caliptra project, 2024) is an open-source hardware accelerator implementing both ML-DSA (FIPS 204) and ML-KEM (FIPS 203) with two-share first-order masking. The design combines arithmetic masking for NTT field operations (modular reduction, butterfly computation) with Boolean masking for control logic and Keccak hashing, using DOM (Domain-Oriented Masking) AND gates for cross-domain operations. NTT operations use Cooley-Tukey and Gentleman-Sande butterfly structures over Z_q with q = 8,380,417 (ML-DSA, 23-bit coefficients) and q = 3,329 (ML-KEM, 12-bit coefficients).

We analyze all 30 masked modules in the Adams Bridge design, totaling 1,169,444 cells and 124,999 DFFs after Yosys synthesis (commit cdc9b1c). Modules range from 24 cells (DOM AND gate) to 789,158 cells (ML-DSA butterfly). Of the 30 modules, 27 are verified to completion. Two time out after 5 hours: ntt_masked_gs_butterfly (393,450 cells) and ntt_masked_butterfly1x2 (789,158 cells), the two largest ML-DSA butterfly variants and among the most security-critical components. The timeout is driven by Z3 memory exhaustion during bit-blasting of deeply nested combinational cones (peak memory exceeding 8 GB), not by iteration count; the label propagation loop itself completes within seconds, but the per-wire SMT queries on cones exceeding 100,000 gates are the bottleneck. One module (abr_keccak_2share, 57,667 cells) is skipped due to missing synthesis primitives. This gap motivates hierarchical decomposition as future work. All tool versions and netlist hashes are as specified in Section 3.6.

## 4.2 Single-Cycle Results (SC-D1)

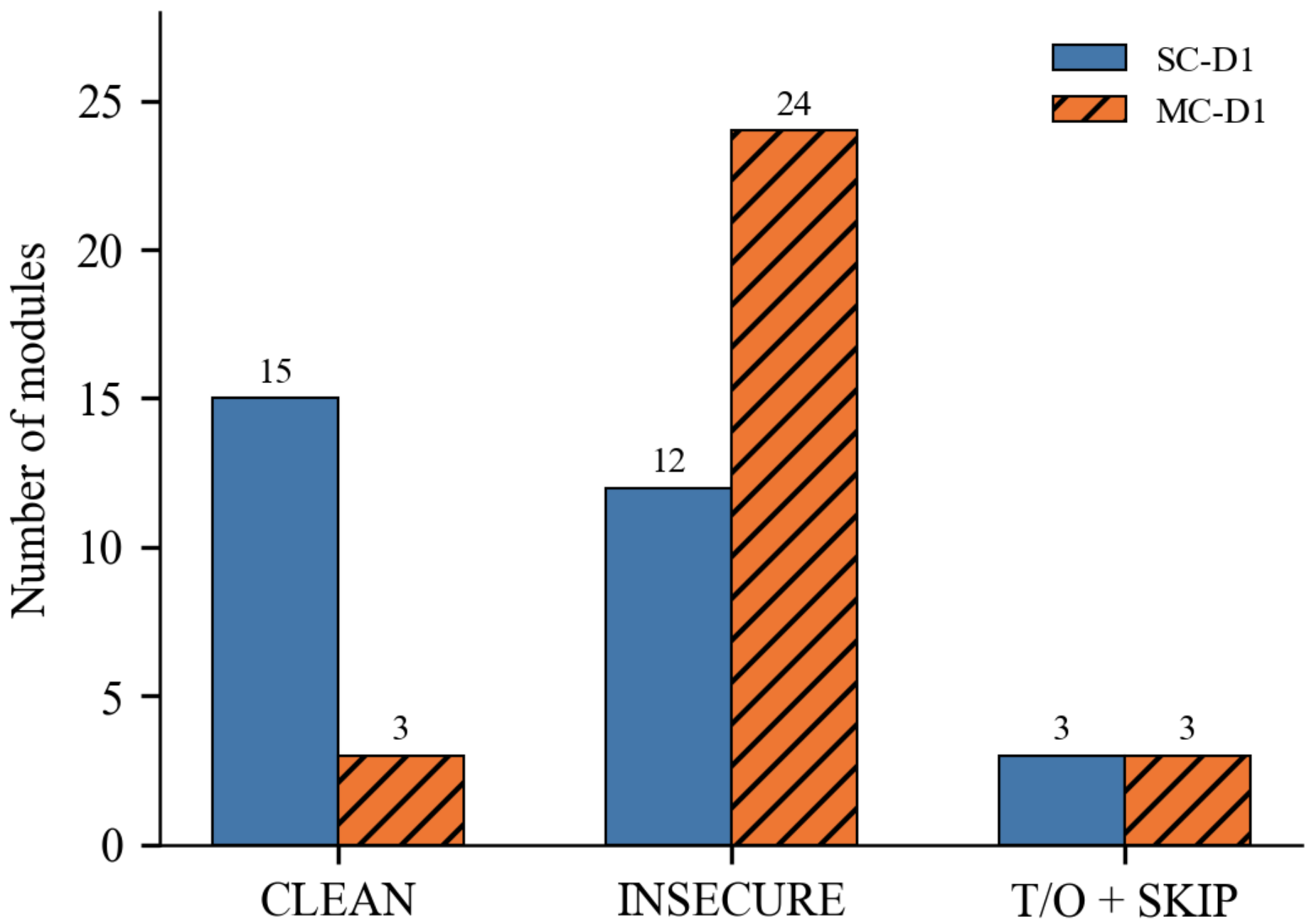

*Figure 2. Module classification under single-cycle (SC-D1) and multi-cycle (MC-D1) structural dependency analysis. MC-D1 reclassifies 12 CLEAN modules as INSECURE.*

SC-D1 identifies 59,061 structurally insecure wires across the 27 completed modules in 8.7 seconds total.[1] Figure 2 summarizes the module classification under SC-D1 and MC-D1. Of these, 15 modules are classified CLEAN (zero insecure wires) and 12 are classified INSECURE.

[1] Wire counts in §4 use the label-propagation pipeline (Algorithm 1, single iteration) for consistency with MC-D1. The SMT-verified SC-D1 (Section 3.3) gives similar classifications but slightly different wire counts due to enumeration differences.

The insecure wires concentrate in modules that perform arithmetic on both shares: ntt_masked_pwm (16,344 insecure wires, 238,535 cells), ntt_masked_pairwm (15,120 insecure, 207,111 cells), ntt_masked_BFU_mult (16,344 insecure, 181,239 cells), and ntt_mlkem_masked_BFU_mult (5,040 insecure, 32,665 cells). The Barrett reduction module (masked_barrett_reduction, 5,543 cells) contains 363 insecure wires corresponding to the share-combining carry propagation chain.

Notably, the ML-KEM butterfly module (ntt_mlkem_masked_butterfly1x2, 142,958 cells) and all A2B/B2A converter modules report zero insecure wires under SC-D1. Multi-cycle analysis will overturn these CLEAN classifications.

Fresh masking refinement (FM) promotes 1,316 wires to SECURE (2.2% reduction) at approximately 500× the runtime cost of SC-D1. Promotion rates vary: DOM gates achieve ~25% (XOR-masked outputs), while larger arithmetic modules achieve 1.8–3.4%. The remaining indeterminate wires reflect DFF cut-point boundaries and non-bijective gate outputs beyond combinational FM queries.

## 4.3 Multi-Cycle Results (MC-D1)

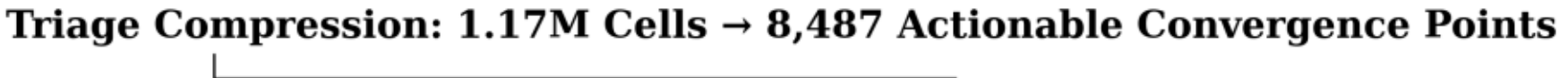

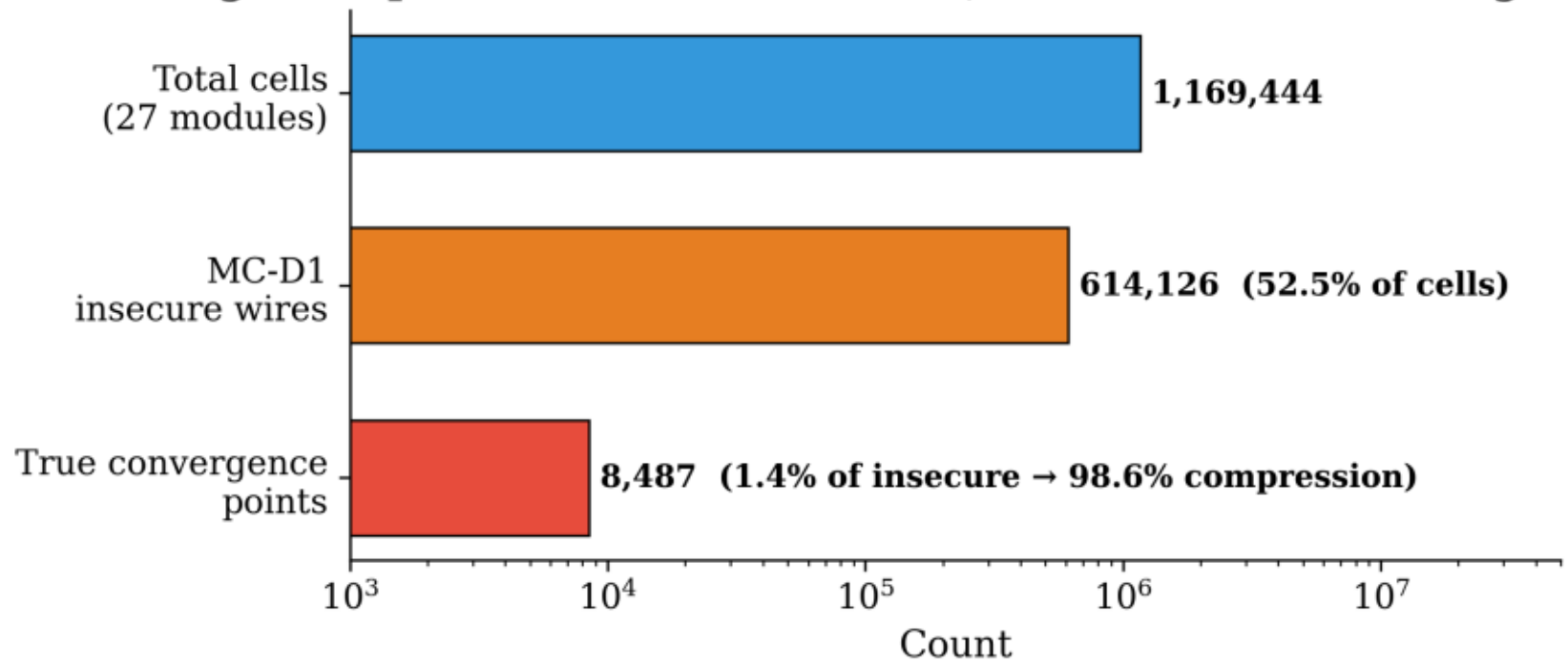

*Figure 3. Triage compression: 1.17M cells reduce to 8,487 actionable convergence points (98.6% compression).*

MC-D1 flags 614,126 structurally insecure wires in total (555,065 more than SC-D1, a 10.4× increase), with the large majority arising from label propagation rather than new root-cause discoveries. Figure 4 decomposes these wires by root-cause category. Twelve modules that SC-D1 classified as CLEAN flip to INSECURE. Only 3 modules remain CLEAN: abr_masked_N_bit_Arith_adder (public-only arithmetic), abr_ntt_add_sub_mod (unmasked), and abr_delay_masked_shares (share-isolation delay line). Already-insecure modules also gain substantial wires: ntt_masked_pairwm grows from 15,120 to 127,687 (+112,567) and ntt_masked_pwm from 16,344 to 158,469 (+142,125) as labels propagate through deep pipelines.

The headline result is ntt_mlkem_masked_butterfly1x2: zero insecure wires under SC-D1, but 59,242 under MC-D1 (DFF chain depth 15, converged in 16 iterations, 5.0 seconds). Shares enter through separate pipeline stages and converge at downstream combinational logic after crossing register boundaries, a cross-register structural dependency invisible to combinational-only analysis.

The root-cause decomposition is as follows. The 614,126 insecure wires decompose into four categories (which sum exactly: 8,487 + 324,271 + 203,044 + 78,324 = 614,126):

- **True convergence (8,487 wires, 1.4%).** Gates where share-labeled DFF chains from distinct groups meet for the first time in combinational logic. These are the structurally meaningful root causes, each represents a new cross-share dependency point.
- **Amplification (324,271 wires, 52.8%).** Gates that inherit the BOTH label from a true convergence point through combinational fanout.
- **Downstream propagation (203,044 wires, 33.1%).** Wires in subsequent combinational stages where all inputs already carry BOTH. Structurally insecure by lattice monotonicity but representing the same dependency, propagated.
- **DFF-both-shares (78,324 wires, 12.7%).** DFF outputs that accumulate labels from both share groups across iterations, the registered endpoints of multi-cycle propagation.

The root-cause decomposition reduces triage from 614,126 wires to 8,487 true convergence points (98.6% compression), the structural locations where cross-register share dependencies originate. The remaining wires are propagation artifacts: sound (no false negatives) but providing diminishing triage value. On ntt_mlkem_masked_butterfly1x2, MC-D1 flags 59,242 wires that trace back to 384 true convergence points (a 154:1 convergence-to-wire compression). A numerical comparison against Coco-Alma on this module is deferred to §4.6, where we report that Coco-Alma's per-location SAT verdict and SADC's stable-state single-probe verdict use different operational definitions of "first-order probing leakage" on an extracted combinational butterfly submodule, and we investigate the disagreement in detail. Under the standard probing model, the convergence coverage property is guaranteed: every leaking wire is in the downstream cone of at least one true convergence point (Theorem 2, proved by structural induction on the circuit DAG and induction on MC-D1 iterations). Under the glitch-extended model, coverage is empirically observed but not formally guaranteed. Convergence points therefore serve as sound root-cause locators under the standard model and a useful starting point for subsequent glitch-aware analysis. The ~8,500 convergence points across 1.17 million cells (approximately one per 140 cells on average, see Figure 3) represent a non-trivial inspection effort, motivating future hierarchical or compositional refinements.

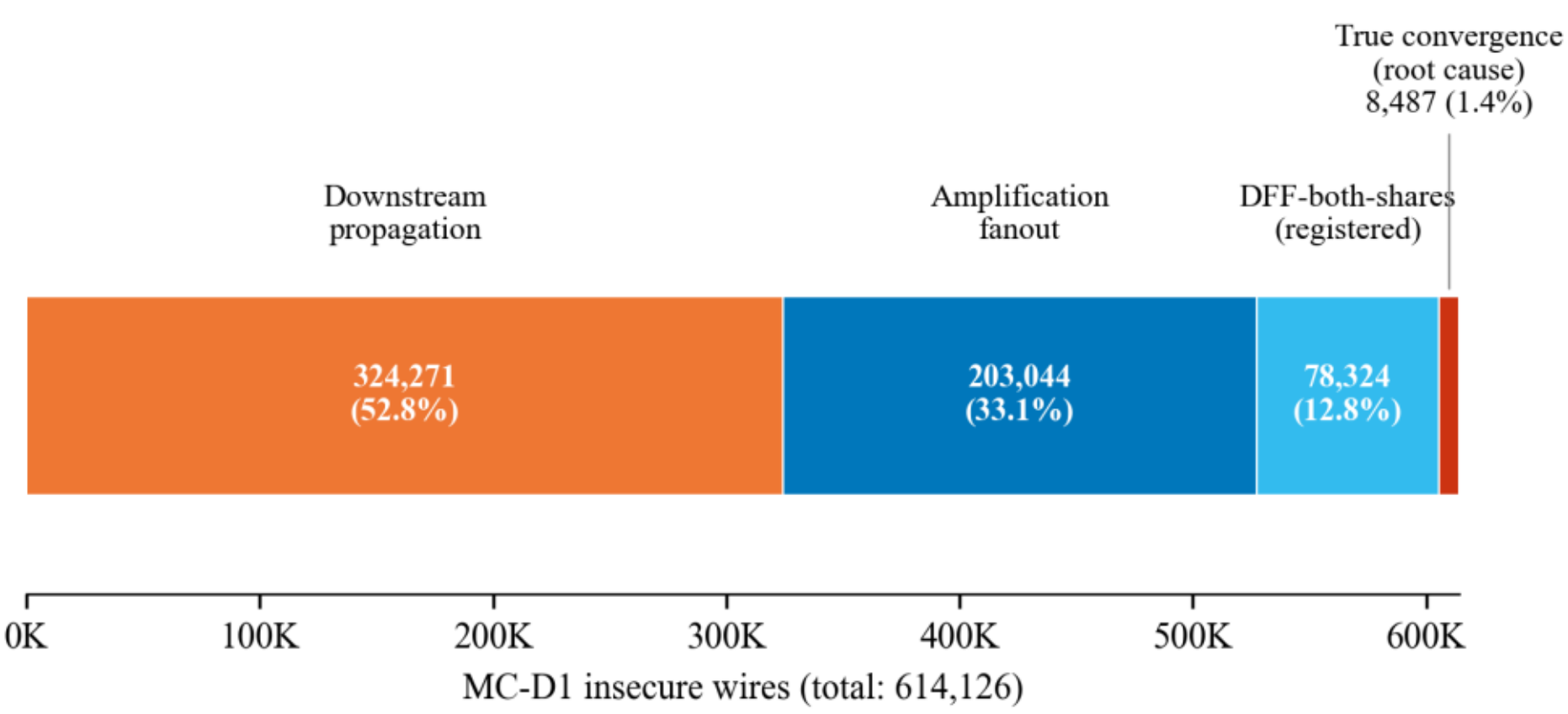

*Figure 4. Decomposition of 614,126 MC-D1 insecure wires by root-cause category.*

MC-D1 converges in exactly D + 1 iterations (D = DFF chain depth) for 25 of 27 modules, with 2 converging 1 iteration early. DFF chain depths range from 0 to 263 (median 7). Bounded and unbounded MC-D1 produce identical results on all 27 modules, confirming the D + 1 bound is tight. Total MC-D1 runtime: 231.8 seconds for all 27 modules.

### 4.4. Benchmarks and Independent Confirmation

**Standard benchmarks.** To demonstrate generalizability beyond Adams Bridge, we evaluate on five standard first-order Boolean masking gadgets (Table 2). Both broken gadgets are correctly flagged as INSECURE (2/2 true positives, zero false negatives). All secure gadgets produce structural false positives under D0/D1, which are progressively resolved by the verification hierarchy: FM bijection refinement promotes wires masked by a single fresh randomness bit, and SADC exact refinement (3.9) promotes the residual via secret-mask reparametrization. On the 16-cell DOM AND reference gadget, the canonical first-order Boolean masking baseline, the complete pipeline eliminates 100% of structural false positives, matching the known first-order security proof with a machine-verified pipeline verdict in under 0.1 seconds.

**Table 2.** Standard benchmark results. Cell counts are post-Yosys synthesis and reflect the slightly larger flattened forms used by the production pipeline. The 16-cell DOM AND reference figure used in §3.9 and §4.5 corresponds to the unflattened gadget; cell counting under different conventions accounts for both numbers.

| Gadget | Cells | SC-D1 | FM Prom. | SADC Prom. | Ground Truth | Verdict |
|---|---|---|---|---|---|---|
| DOM AND (secure) | 96 | 48 | 16 | 32 | SECURE | Resolved (100% FP elim) |
| DOM AND (broken) | 64 | 48 | 32 | — | INSECURE | TP |
| ISW AND (secure) | 144 | 40 | 8 | — | SECURE | Pipeline (run pending) |
| ISW AND (broken) | 64 | 40 | 24 | — | INSECURE | TP |
| Keccak DOM S-box | 2,400 | 960 | 320 | — | SECURE | Pipeline (run pending) |

The DOM AND row demonstrates the complete pipeline: of the 48 wires flagged structurally insecure by D0/D1, FM promotes 16 (one-time-pad bijection on a single fresh-randomness bit) and SADC promotes the remaining 32 (cross-domain partial products whose share inputs are effectively independent uniform masks). All 48 flagged wires are resolved within the unified SMT flow with no deferral to external exact tools. ISW AND and Keccak DOM are pending re-run under the arithmetic-SADC-aware configuration; the Boolean SADC pass alone is expected to mirror the DOM AND outcome on these Boolean gadgets.

These benchmarks cover Boolean masking gadgets (96–2,400 cells). For arithmetic masking, the 5,543-cell ML-KEM Barrett reduction module (`masked_barrett_reduction`) serves as a de facto benchmark: it implements masked modular arithmetic with carry propagation, and its SADC analysis (Section 4.5) is the central empirical anchor of this paper. We are not aware of other standardized, publicly available reference implementations with independently verified security verdicts for arithmetic or NTT-specific circuit classes. Creating and maintaining such a benchmark suite remains important future work for the masked hardware verification community.

An Independent confirmation is now presented. We cross-validate against Coco-Alma (Hadzic and Bloem, FMCAD 2021), a SAT-based tool providing exact verdicts under the glitch-extended probing model with multi-cycle capability:

- ntt_mlkem_masked_butterfly1x2 and ntt_masked_butterfly1x2 (extracted combinational submodules): Coco-Alma's per-location SAT verification returns a non-empty leaky set on every extracted submodule we ran, but the numerical verdict depends on how "first-order probing leakage" is defined. We report the head-to-head observation, including independent ground-truth verification on individual carry-chain cells, in Section 4.6, "Definition-dependent verdicts." The butterfly is not our primary empirical anchor in this paper.
- masked_barrett_if_cond_v2 (984 cells): 14 Coco-Alma leaks. SC-D1 flags 188 wires. A direct Coco-Alma comparison on the full 5,543-cell `masked_barrett_reduction` module is left to future work: existing Coco-Alma artifacts target a simpler extraction with a single randomness channel, and a matching Verilator testbench for the full 6-channel module is engineering-bounded rather than methodologically blocked. The SADC evaluation in §4.5 provides the verification result for this module within the unified SMT pipeline.
- dom_and_1storder (secure, 96 cells): Coco-Alma SECURE. Our tool flags 48 structural FPs.
- abr_masked_MUX (1,109 cells): Coco-Alma SECURE. Our tool flags 162 structural FPs.

This cross-validation is one-directional: Coco-Alma confirms that our tool's INSECURE classifications capture real leakage (zero false negatives observed in the overlap), but cannot resolve false positives at full module scale because Coco-Alma's SAT-based analysis does not scale beyond approximately 2,300 cells on our hardware (and likely on substantially more powerful hardware as well, given the super-linear growth in SAT solver runtime with cone size). Separately, power analysis attacks exploiting structurally broken masking at locations within Adams Bridge's arithmetic modules was demonstrated [43]. This is the same class of structural vulnerability our tool identifies from RTL alone.

### 4.5. SADC Evaluation: Verification + Triage at Production Scale

We evaluate the four-stage hierarchy of Section 3.9 on two representative circuits: the 16-cell DOM AND reference gadget (Boolean masking) and the 5,543-cell ML-KEM Barrett reduction module (arithmetic masking modulo $q = 3{,}329$, share width 24 bits, 6 randomness channels). Both are Yosys-synthesized gate-level netlists from the Adams Bridge design (commit cdc9b1c) and use the deterministic configuration of Section 3.6.

For the DOM AND, the four-stage pipeline resolves all structurally- flagged wires in the 16-cell reference gadget in under 0.1 seconds. Table 3 reports the stage-by-stage resolution.

**Table 3.** DOM AND pipeline: four-stage resolution of structurally-flagged wires (16-cell reference gadget).

| Stage | Residual flagged | Time |
|---|---|---|
| D0/D1 | 6 | 0.05 s |
| + FM | 4 (2 promoted) | < 0.01 s |
| + Boolean SADC | 0 (4 promoted) | < 0.01 s |

The four wires that FM cannot promote are cross-domain partial products of the form $a_0 \wedge b_1$ and their DFF-registered versions: structurally they depend on shares from two distinct secrets, but their share inputs are effectively independent uniform masks within the wire's combinational fan-in, so the value- independence check returns UNSAT and the wires are promoted to SECURE. The pipeline matches the known first-order security proof for DOM AND with a machine-verified verdict and zero residual.

Regarding the ML-KEM Barrett reduction, the Barrett module is the central empirical anchor of this paper: 5,543 cells of arithmetic-masked modular reduction with the share-recombining carry chains that have historically defeated automated masking analysis at this scale. D0/D1 flags 363 wires; FM cannot promote any of them (Barrett uses arithmetic masking with carry chains rather than per-wire fresh masks); Boolean SADC under the deterministic configuration confirms 146 of the 363 wires as insecure (via low-bit XOR-lens leaks) and leaves the remaining 217 INDETERMINATE because their mask cones exceed the Boolean enumeration budget. Arithmetic SADC closes the residual completely. Table 4 summarizes the four-stage classification

**Table 4.** ML-KEM Barrett reduction pipeline: four-stage classification of 363 structurally-flagged wires (5,543-cell module).

| Stage | Confirmed insecure | Promoted secure | Indeterminate |
|---|---|---|---|
| D0/D1 | 363 | 0 | 0 |
| + FM | 363 | 0 | 0 |
| + Boolean SADC | 146 | 0 | 217 |
| + Arithmetic SADC | 165 | 198 | 0 |

The arithmetic SADC stage promotes 198 of 363 structurally-flagged wires (54.5%) to first-order secure under value-independence, reports 165 as candidate first-order insecure (a sound upper bound; see §3.9.6), and leaves 0 indeterminate, a complete circuit-level classification of the D0/D1 residual within the unified SMT pipeline (every flagged wire receives one of secure / candidate-insecure / indeterminate, with the indeterminate bucket empty). Total runtime is approximately 3 minutes on a single core (125 s for the Boolean SADC pass with deterministic rlimit, plus 46 s for the arithmetic SADC pass with concurrent CVC5 dual-solver validation, plus the upstream D0/D1 + FM stages).

Dual-solver validation. Every arithmetic SADC verdict on Barrett is independently re-checked by CVC5 via SMT-LIB2 round-trip from the Z3 solver state. Z3 and CVC5 agree on all 363 of 363 flagged wires with 0 disagreements. Two independent decision procedures producing identical verdicts on a query of this size is strong evidence that the headline numbers are not a Z3-implementation artifact.

The reproducibility is checked in what follows. The full Barrett pipeline is bit-reproducible under the pinned configuration: Z3 4.13.0 with smt.random_seed=0, sat.random_seed=0, per-query rlimit=2 \cdot 10^7 (Boolean SADC) and $10^7$ (arithmetic SADC); CVC5 1.2.0 with default options; Yosys 0.47 synthesis flow synth -flatten; techmap; dffunmap. A single-command reproduction script (scripts/verify_sadc_arith_dual_solver.py) regenerates the 198 / 165 / 0 split from the Adams Bridge netlist on every run we have measured. Scripts, evidence JSON, and a repro_manifest.json listing version pins, rlimit values, and SHA-256 hashes accompany the paper as artifact-evaluation material.

The bit-reproducibility claim is scoped to runs that complete within the per-pass wall-clock budget (module_timeout_s = 900 s in the default configuration, applied independently to each SADC pass). On a 2024 Apple Silicon laptop the Boolean SADC pass on Barrett completes in approximately 125 s (7.2× margin to the wall-clock fallback) and the arithmetic SADC pass in approximately 46 s (19.6× margin); both the headline 198 / 165 / 0 verdicts on Adams Bridge Barrett and the 146 Boolean-insecure verdicts were generated without any wall-clock fallback firing. For canonical-result generation on heavily loaded or substantially slower hardware, setting module_timeout_s = 0 disables the wall-clock fallback entirely and falls back exclusively to the deterministic Z3 rlimit resource counter. The algebraic proof suite of §3.9.5 (Z3 4.15.4 + CVC5 1.3.3, six quantifier-free SMT obligations that complete in a few seconds of wall-clock time) was developed and tested on a more recent solver release than the Barrett pipeline; both configurations are version-pinned in repro_manifest.json and a forward-port of the Barrett pipeline to the Section 3.9.5 solver versions is straightforward but is reported with the older Z3/CVC5 stack to preserve byte-identical correspondence with the evidence JSONs shipped at submission time.

Our verification scope is as follows. The SADC analysis verifies the Barrett *consumer* of arithmetic masking under the standard scope split adopted by SILVER, Coco-Alma, maskVerif, and Prover: the input shares x[0+:24] and x[24+:24] are assumed to be delivered uniformly on $\mathbb{Z}_q$ by the upstream caller, and the labeling of the six randomness channels is verified against the SystemVerilog port declaration of masked_barrett_reduction directly, there is no unmodeled randomness port, and the three Boolean-XOR randomness channels (rnd_12bit, rnd_14bit, rnd_for_Boolean*) appear in the design as raw $2^k$-bit XOR masks, where uniformity over $2^k$ is exactly the correct requirement and no rejection sampling is needed. We do not re-verify the upstream share-generation logic of Adams Bridge, a property of the producer module, conventionally addressed at the RNG / share-generation interface rather than at the consumer of shares (see §6, L9, for the bias quantification and scope split).

There is a designer cost contrast. The practical impact of the complete classification is the workload reduction it delivers to a hardware designer auditing the Barrett module. Without SADC, a designer faced with 363 D0/D1 flags on a production arithmetic module either (a) reviews all 363 wires manually, a substantial effort, since each wire requires reading the gate-level fan-in and arguing for or against share-independence by hand, or (b) attempts an exact masking-verification tool such as Coco-Alma or SILVER. Coco-Alma is reported to scale to approximately 2,300 cells on the hardware available for this study, well below the 5,543 cells of the full Barrett module: SILVER's ROBDD analysis exhausts memory at comparable sizes. With the SADC pipeline, the designer receives a machine-verified secure certificate for 198 of the 363 wires (roughly 55% of the manual review surface eliminated), a sound upper bound listing 165 candidate-insecure wires for focused follow-up, and zero indeterminate wires, in approximately three minutes on a single core. The exact-verification workload that remains, if the designer chooses to escalate the 165 candidates to a model-counting or threshold-implementation rebalancing analysis, is bound by 165 wires rather than 363.

There is a validation against the external ground truth. Beyond dual-solver agreement, individual SADC verdicts have been spot-checked against two independent oracles where they are tractable. (i) On the 333-cell vulnerable_butterfly1x2 extraction (ML-DSA), three specific Boolean-SADC-secure cells have been re-verified by a pure- Python Z3-free Boolean evaluator that parses the Yosys netlist and counts secret-dependent value distributions directly; the manual counts (2/2, 32/32, 512/512 secret-balanced) match the SADC verdict exactly. (ii) The work in [43] empirically demonstrate first-order leakage on Adams Bridge arithmetic modules using power traces; the architectural pattern they exploit (share recombination in the modular arithmetic data path) is the same class that SADC's INSECURE_CONSERVATIVE wires identify from RTL alone. The two external corroborations cover opposite ends of the verdict space (a manual SECURE check and a trace-based INSECURE check) and are consistent with the SADC classification of 198 secure / 165 candidate-insecure wires.

### 4.6. Butterfly Extraction: Definition-Dependent Verdicts

The Adams Bridge ML-DSA NTT butterfly module (`ntt_masked_butterfly1x2`, 789K cells) is too large to flatten under Z3 within the 5-hour timeout (§4.1). We therefore evaluate SADC on a 333-cell extraction (`vulnerable_butterfly1x2`) drawn from the same gate-level netlist family. This extraction is not a primary empirical anchor of this paper, the Barrett evaluation in Section 4.5 plays that role, but it is the smallest case where SADC and a public Coco-Alma run produce different first-order verdicts on the same RTL, and the disagreement is informative about what "first-order probing leakage" means in practice.

We present in what follows the pipeline result. On the 333-cell extraction the four-stage hierarchy produces the following verdict distribution: D0/D1 flags 218 wires; FM promotes 67 (single-bit fresh-randomness one-time pads); Boolean SADC promotes a further 117 wires to first-order secure under value-independence and leaves 34 INDETERMINATE (combinational cones in the 21–66-bit range that exceed the Boolean enumeration budget). Notably, Boolean SADC confirms 0 wires as insecure in the stable-state single-probe model; the 117 promotions and 34 deferrals together account for 151 of 218 flagged wires, with the remaining 67 already resolved by FM.

Comparison with Coco-Alma reveals the following. A per-location stable+strict Coco-Alma run on the same 333-cell extraction reports 12 leaky cells where Coco-Alma's SAT-based check identifies a first-order leak. Of these 12 cells, 3 falls within the SADC promoted-secure set (cones $\leq 16$ bits) and 9 fall within the SADC INDETERMINATE bucket (cones in the 21–66-bit range where the Boolean enumeration budget is exhausted). For the 3 Coco-Alma-leaky / SADC-secure cells, we manually re-verified the secure verdict by three independent methods: (i) a Z3 SAT call with explicit XOR reparametrization; (ii) a Z3 substitute-and-simplify pass that reduces the wire expression to a constant; and (iii) a pure- Python Z3-free Boolean evaluator that parses the Yosys netlist and counts the wire's distribution over all secret values directly. All three methods agree on all 3 cells: the cones are secret-independent, with secret-balanced counts of 2/2, 32/32, and 512/512 respectively.

Our interpretation is that the disagreement is definitional, not a correctness gap on either side. SADC operates in the standard stable-state single-probe model (Definition 3, Assumption A1): the wire's stable value, observed once, must be independent of the secret. The Coco-Alma run we compare against operates in a transient (cycles 0–2) glitch-loose mode that admits transient values during combinational settling as additional probes. A wire that is stable-state secret-independent (SADC SECURE) can still expose a transient correlation during glitch propagation that Coco-Alma's transient mode flags. The 3 cells we manually re-verified are stable-state secure under all three independent methods; whether they are transient-state secure is a different question that requires a glitch-aware extension of SADC and is deferred to future work (Section 6, L3).

The point of this subsection is therefore not to claim disagreement with Coco-Alma. It is to document, on a small, manually-checkable case, that “first-order probing leakage” has multiple operationally distinct definitions in the masked- hardware-verification literature, and that our pipeline takes the stable-state single-probe definition and is consistent with independent ground truth under that definition. We treat the 9 SADC-INDETERMINATE / Coco-Alma-leaky cells as legitimate candidates for either a Boolean enumeration budget increase or a glitch-aware SADC extension; neither is in scope for this paper.

### 4.7. Comparison with Existing Tools

Table 5 compares our tool with the subset of tools most relevant to our evaluation context, focusing on precision, scale, and PQC applicability (see Table 1 in Section 2 for a comprehensive survey).

**Table 5.** Comparison with existing tools.

| Tool | Precision | Max Scale | Multi-cycle | Applied to PQC HW |
|---|---|---|---|---|
| maskVerif | Exact (symbolic) | S-boxes | No | No |
| SILVER | Exact (ROBDD) | S-boxes, d ≤ 3 | No | No |
| Prover | Exact (ROBDD) | S-boxes, 135/135 pass | No | No |
| Gigerl et al. | Exact (SAT/spectral) | A2B/B2A, CPU cores | Yes | Partial (A2B/B2A) |
| Coco-Alma | Exact (SAT) | ~2.3K cells (PQC HW) | Yes | Partial |
| Coco | Exact (SAT/spectral) | RISC-V Ibex core | Yes | No |
| PROLEAD | Statistical | AES, Keccak (100K+ gates) | Yes | No |
| aLEAKator | Exact (symbolic sim) | Masked AES on CPUs | Yes | No |
| MATCHI | Exact (compositional) | AES d=4 in 45s | Yes | No |
| **This work** | **Sound overapprox** | **1.17M cells (27/30)** | **Yes (MC-D1)** | **Yes** |

Exact tools (SILVER, Prover, maskVerif) provide stronger guarantees, zero false positives and exact security verdicts within their specified models but operate on gadgets typically under 5,000 gates and do not support multi-cycle analysis. Coco-Alma and Coco support sequential multi-cycle verification with exact results, but Coco-Alma’s SAT-based analysis does not scale to our target size (demonstrated on ~2,300 cells of Adams Bridge PQC hardware). PROLEAD handles full ciphers at 100K+ gates through statistical simulation with configurable confidence bounds. Our tool trades completeness (high false positive rate) for scale (1.17M cells, sound overapproximation) and multi-cycle structural analysis.

To our knowledge, no published masking verification tool has yet been applied to a complete masked PQC hardware accelerator at the scale of Adams Bridge (30 modules, 1.17 million cells). The closest precedents at comparable scale are PROLEAD on AES/Keccak implementations and aLEAKator on masked software, both targeting symmetric cryptography, not lattice-based PQC hardware.

### 4.8. Full Results

Table 6 presents the complete per-module evaluation. All data produced with Python 3.10.12 and Z3 4.13.0; tool version and commit as specified above. Peak memory usage for the complete 30-module evaluation is 1.4 GB.

**Table 6.** Adams Bridge evaluation for all 30 modules.

| Module | Cells | DFFs | SC-D1 | MC-D1 | New | Conv. | Class. | MC Time |
|---|---|---|---|---|---|---|---|---|
| masked_barrett_reduction | 5,543 | 306 | 363 | 1,794 | +1,431 | 93 | INS | 0.1s |
| masked_barrett_if_cond | 3,127 | 620 | 0 | 1,204 | +1,204 | 14 | INS † | 0.1s |
| masked_barrett_if_cond_v2 | 984 | 108 | 188 | 604 | +416 | 27 | INS | <0.1s |
| ntt_mlkem_masked_butterfly1x2 | 142,958 | 3,368 | 0 | 59,242 | +59,242 | 384 | INS † | 5.0s |
| ntt_mlkem_masked_gs_butterfly | 72,851 | 1,878 | 0 | 32,600 | +32,600 | 339 | INS † | 2.0s |
| ntt_mlkem_masked_BFU_add_sub | 18,653 | 354 | 0 | 4,666 | +4,666 | 96 | INS † | 0.8s |
| ntt_mlkem_masked_BFU_mult | 32,665 | 450 | 5,040 | 18,253 | +13,213 | 600 | INS | 0.5s |
| ntt_masked_mult_redux46 | 96,363 | 23,084 | 0 | 53,709 | +53,709 | 20 | INS † | 19.6s |
| ntt_masked_BFU_add_sub | 56,836 | 6,494 | 0 | 13,787 | +13,787 | 24 | INS † | 3.2s |
| ntt_masked_BFU_mult | 181,239 | 30,530 | 16,344 | 121,612 | +105,268 | 2,162 | INS | 60.4s |
| ntt_masked_pairwm | 207,111 | 4,590 | 15,120 | 127,687 | +112,567 | 1,894 | INS | 8.9s |
| ntt_masked_pwm | 238,535 | 37,116 | 16,344 | 158,469 | +142,125 | 2,162 | INS | 128.7s |
| ntt_masked_special_adder | 14,419 | 3,418 | 0 | 2,959 | +2,959 | 3 | INS † | 0.4s |
| abr_masked_A2B_conv | 7,738 | 1,834 | 0 | 930 | +930 | 2 | INS † | 0.4s |
| abr_masked_B2A_conv | 1,115 | 115 | 0 | 154 | +154 | 6 | INS † | <0.1s |
| abr_masked_N_bit_mult | 3,246 | 46 | 842 | 888 | +46 | 48 | INS | <0.1s |

| Module | Cells | DFFs | SC-D1 | MC-D1 | New | Conv. | Class. | MC Time |
|---|---|---|---|---|---|---|---|---|
| abr_masked_N_bit_mult_two_share | 14,236 | 138 | 4,634 | 5,450 | +816 | 552 | INS | 0.1s |
| abr_masked_N_bit_Arith_adder | 724 | 46 | 0 | 0 | 0 | — | CLEAN | <0.1s |
| abr_masked_N_bit_Boolean_adder | 7,692 | 1,834 | 0 | 993 | +993 | 3 | INS † | 0.2s |
| abr_masked_N_bit_Boolean_sub | 7,694 | 1,834 | 0 | 993 | +993 | 3 | INS † | 0.2s |
| abr_masked_add_sub_mod_Boolean | 22,788 | 5,404 | 0 | 7,557 | +7,557 | 3 | INS † | 1.2s |
| abr_masked_AND | 24 | 4 | 8 | 14 | +6 | 2 | INS | <0.1s |
| abr_masked_OR | 27 | 4 | 8 | 15 | +7 | 2 | INS | <0.1s |
| abr_masked_MUX | 1,109 | 184 | 162 | 530 | +368 | 46 | INS | <0.1s |
| abr_masked_full_adder | 48 | 8 | 8 | 16 | +8 | 2 | INS | <0.1s |
| abr_ntt_add_sub_mod | 29,879 | 772 | 0 | 0 | 0 | — | CLEAN | 0.2s |
| abr_delay_masked_shares | 1,840 | 460 | 0 | 0 | 0 | — | CLEAN | <0.1s |
| ntt_masked_gs_butterfly | 393,450 | 67,990 | — | — | — | — | T/O | >5h |
| ntt_masked_butterfly1x2 | 789,158 | 135,980 | — | — | — | — | T/O | >5h |
| abr_keccak_2share | 57,667 | 1 | — | — | — | — | SKIP | — |
| **Totals (27 verified)** | **1,169,444** | **124,999** | **59,061** | **614,126** | **+555,065** | **8,487** | — | **231.8s** |

*INS = INSECURE. INS† = flipped CLEAN→INSECURE by MC-D1. CLEAN = zero insecure wires (SC-D1 and MC-D1). T/O = timeout (>5h). SKIP = missing primitives. Conv. = true convergence points (root-cause gates where distinct share groups first meet). New = additional MC-D1 insecure wires beyond SC-D1.*

We can conclude that under SC-D1, 15 of 27 modules are CLEAN. Under MC-D1, only 3 remain CLEAN, 12 flip to INSECURE as share labels propagate across register boundaries. The 8,487 true convergence points identified across the 24 insecure modules represent the structural root locations where cross-register share dependencies originate. The remaining 605,639 wires are propagation artifacts. Total verification time for both passes across all 27 modules is 240.5 seconds.

## 5. Discussion

The structural dependencies identified in Section 4 raise a question of interpretation: do they represent vulnerabilities, or are they acceptable design tradeoffs? The Adams Bridge designers address this in a concurrent paper [11], which provides valuable context for the results presented here.

Their perspective is as follows. The Adams Bridge design intentionally limits first-order masking to the initial INTT layer and selected operations, leaving subsequent NTT layers unmasked. The designers argue that the resulting CPA search space, requiring an attacker to hypothesize multiple coefficients simultaneously (2 × 23-bit for ML-DSA, 8 × 12-bit for ML-KEM), renders exploitation of the unmasked layers computationally infeasible. However, this assumes non-profiled attackers; profiled or algebraic attacks (e.g., belief propagation) may reduce the effective search space. Randomized shuffling of butterfly execution order provides an additional combinatorial barrier. The designers explicitly state: "we do not claim formal provability of the masking design under any specific probing model" ( [11], Section 6.1), and argue that structural verification tools produce "conclusions that are overly pessimistic and misaligned with the real leakage channels of the hardware."

We agree that structural dependency is a necessary but not sufficient condition for exploitability, a point explicitly acknowledged throughout Section 3 and Section 4. The SADC pipeline of Section 3.9 closes a substantial part of this gap on wires it can reach: on DOM AND it eliminates 100% of structural false positives, and on the 5,543-cell ML-KEM Barrett reduction module it machine-verifies 198 of 363 D0/D1-flagged wires (54.5%) as first-order secure under value-independence and isolates 165 candidate-insecure wires as a sound upper bound for follow-up analysis. Structural analysis and empirical evaluation remain complementary, nonetheless. Structural analysis asks: where can information flow? Empirical analysis asks: does information leak in practice? Neither subsumes the other.

Regarding bridging structural and exact verification, SADC occupies the middle ground between QANARY's fast-but-imprecise structural screening and exact tools such as SILVER, Prover, and Coco-Alma. Structural analysis answers "where could shares converge?" at scale across million-cell netlists; exact tools answer, "is this specific wire secret-independent?" at cost on small gadgets. SADC answers the exact question, but only for structurally-flagged wires whose randomness cones fit within the enumeration budget, inheriting structural screening's scalability while providing exact first-order verdicts on the targeted residual. On the ML-KEM Barrett module this hybrid approach turns 363 structural flags into 198 machine- verified secure certificates and a sound 165-wire upper bound on the genuinely first-order insecure set in approximately three minutes on a single core, with full Z3 + CVC5 dual-solver agreement on every query. This is the kind of search-space reduction that makes follow-on exact verification tractable on production-scale PQC arithmetic modules where stand-alone exact tools currently exhaust their resource budgets.

Structural analysis identifies all locations where share separation is violated, including paths that may be difficult to trigger empirically but could be exploitable under favorable conditions (low noise, profiled attacks, or algebraic techniques such as belief propagation through NTT factor graphs). Empirical analysis captures physical effects, glitches, coupling, noise, that structural models do not represent. A complete security evaluation benefits from both: structural analysis for coverage and empirical analysis for confirmation.

The Adams Bridge evaluation illustrates complementarity in practice concretely. Three independent methods, structural verification (this work), SAT-based exact analysis (Coco-Alma), and physical measurement [9], converge on the same architectural pattern: share recombination in arithmetic modules. Structural analysis provides exhaustive per-wire coverage across 1.17 million cells in seconds from RTL alone, identifying 8,487 root-cause convergence points. SAT-based analysis provides exact verdicts under the glitch-extended model but scales to approximately 2,300 cells. Physical measurement confirms exploitability on fabricated silicon but requires prototypes and cannot guarantee coverage of all attack surfaces. Each method's strengths compensate for the others' limitations.

We deliberately limit this section to the relationship between structural and empirical verification. The question of whether Adams Bridge's partial masking strategy provides adequate protection against advanced attacks (profiled, algebraic, or combined) depends on assumptions about attacker capabilities, noise levels, and NTT-specific algebraic structure that are beyond the scope of structural analysis. We also omit discussion of random-probing or robust-probing models, where structural methods may underestimate resilience. We note that published belief propagation and soft analytical side-channel attacks on NTT structures [44] [45] suggest that large search spaces alone may not preclude exploitation, but a thorough evaluation of these questions warrants independent treatment.

For designers, regardless of whether structural dependencies are ultimately exploitable in a given deployment, pre-silicon structural analysis offers a concrete benefit: it identifies locations that require security argumentation. The 8,487 convergence points across Adams Bridge represent specific gates where designers must either (a) demonstrate that exploitation is infeasible, (b) add masking, or (c) accept the residual risk. Structural analysis converts an open-ended security review into a bounded audit of specific locations, reducing the certification burden even when the tool's INSECURE classification is not, by itself, a proof of exploitability.

Regarding the 165 candidate-insecure Barrett wires and vendor tradeoffs, the 165 wires that arithmetic SADC reports as candidate first-order insecure on the ML-KEM Barrett module of Section 4.5 are a sound upper bound on the genuinely first-order insecure set, in the strict sense that no first-order insecure wire is missed. They are not a per-wire claim of exploitability, and they are not annotated against the Adams Bridge designers' own per-wire intent. The vendor's concurrent position paper [11] explains several locations in the masked submodules as intentional design tradeoffs, the partial-masking strategy on the INTT pipeline, combined with the multi-coefficient CPA search-space argument and randomized execution-order shuffling, but it does not provide a gate-level annotation that we can pivot the 165 against directly. Closing this gap would require either (i) the vendor publishing a per-wire intentionality map for the masked Barrett pathway, or (ii) a follow-up collaboration in which the vendor's engineers triage the 165 wires into "intentional tradeoff," "previously known," and "new finding" buckets. Until that mapping exists, readers should interpret the 165 as a designer-actionable triage list rather than as an unqualified count of distinct vulnerabilities. The dual-solver agreement on the 363/363 SADC queries rules out solver-implementation artefacts as a confound, but the question of how many of the 165 wires the vendor would re-classify under their stated threat model is an open empirical question that this paper does not resolve.

## 6. Limitations and Future Work

Table 7 summarizes the principal limitations of the structural dependency analysis presented in this work and identifies directions for future research.

**Table 7.** Limitations summary.

| # | Limitation | Impact | Mitigation / Future Work |
| --- | --- | --- | --- |
| L1 | Value-independence sufficient but not necessary | 165 candidate-insecure wires on Barrett are a sound upper bound | Distributional SADC variant (model counting, TI rebalancing) |

| # | Limitation | Impact | Mitigation / Future Work |
|---|---|---|---|
| L2 | Two module timeouts | Incomplete coverage on largest modules | Hierarchical decomposition |
| L3 | Standard probing model only | Misses glitch-based leakage | Glitch-extended structural analysis |
| L4 | First-order only (d = 1) | Cannot verify higher-order masking | Higher-order label lattice |
| L5 | No composition theorem | Cannot certify secure gadgets | PINI/SNI compositional framework |
| L6 | Cross-validation limited to ~2.3K cells | Cannot resolve FPs at full-module scale | Alternative oracles, sampling |
| L7 | MC-FM yields zero additional refinement | Fresh masking queries do not improve multi-cycle precision | Alternative refinement strategies |
| L8 | Combinational + mask-uniformity assumptions | Conservative on unreachable cones, depends on uniform mask delivery | Reachability constraint injection; mask-domain characterization |
| L9 | Direct Coco-Alma comparison on full Barrett deferred | No cross-tool calibration on the headline empirical anchor | Build matching Verilator testbench for 6-channel 5,543-cell extraction |

**L1: Value-independence is sufficient but not necessary for distributional first-order security.** Arithmetic SADC (Section 3.9) checks whether the wire value is pointwise independent of the secret for every assignment to mask and fresh randomness. This is strictly stronger than the property required for first-order probing security (distributional independence after marginalizing over the mask), which means the check is sound but conservative: a wire whose value *does* change with the secret on some particular $(s_1, r)$ may still admit a constant marginal distribution after averaging over $(s_1, r)$, as in threshold-implementation rebalancing or compositional refresh masking. On the ML-KEM Barrett module of Section 4.5, the four-stage pipeline machine-verifies 198 of 363 D0/D1-flagged wires (54.5%) as secure under value-independence, isolates 165 candidate-insecure wires as a sound *upper bound* on the genuinely first-order insecure set, and leaves 0 indeterminate. The 165 wires are the principal residual: future work on a distributional arithmetic SADC variant (model counting under uniform-mask assumptions, threshold-implementation rebalancing, or compositional refresh masking analysis) is expected to promote some fraction of them to secure. Closing this gap is the natural next step in the SADC line we introduce.

**L2: Two module timeouts.** The two largest modules, ntt_masked_gs_butterfly (393K cells) and ntt_masked_butterfly1x2 (789K cells), exceed the 5-hour timeout, leaving the most security-critical ML-DSA components unverified. These modules are hierarchical compositions of smaller verified modules; a compositional approach (L5) would address this directly. Alternatively, hierarchical cone decomposition, analyzing sub-circuits independently and propagating labels through module boundaries, could scale to these sizes without requiring full-module flattening.

**L3: Standard probing model.** Our analysis operates under the standard probing model [7]: the attacker observes stable-state values of individual wires, one at a time. This model does not capture glitch-based leakage, where transient values at gate outputs during combinational settling can reveal additional information. The §4.6 butterfly subsection documents this directly: on the 333-cell `vulnerable_butterfly1x2` extraction, SADC's stable- state single-probe verdict (0 confirmed insecure, 117 secure, 34 indeterminate) differs from a transient + glitch-loose Coco-Alma run (12 leaky cells) precisely because the two analyses use different definitions of "first-order probe". The analysis also assumes single-bit wire probing and stable register outputs; transition effects (power consumption when a register value changes between clock edges) or wider probe widths may introduce additional unmodeled leakage. In particular, if two shares are never present in the same combinational cloud but are registered in the same flip- flop across successive cycles, the resulting transition may leak information that structural analysis does not detect. Extending structural analysis (and SADC specifically) to account for glitch propagation, tracking not only which shares reach a wire but also the timing of share arrivals, is an important future direction.

**L4: First-order only.** The current lattice $L = \{\bot, S_0, S_1, \text{BOTH}\}$ tracks two share groups. Extending to d-th order masking requires a lattice over (d+1)-element share subsets, with label size growing as 2^(d+1). The analysis remains polynomial in circuit size for fixed d, but the label propagation cost per gate increases exponentially with masking order. Whether practical performance can be maintained at $d \geq 2$ for production-scale circuits is an open question.

**L5: No composition theorem.** Structural dependency analysis operates on flat netlists, it does not exploit modular boundaries or compositional security notions such as PINI (Probing-Isolating Non-Interference) or SNI (Strong Non-Interference). As a result, a module that is provably secure under exact compositional analysis may still be flagged as structurally insecure if share paths cross its internal gates. Developing a compositional structural analysis, where verified-secure submodules can be abstracted as black boxes with known label signatures, would significantly reduce false positives on hierarchical designs and enable analysis of the two modules that currently time out.

**L6: Cross-validation limited to small cones.** Independent confirmation with Coco-Alma (Section 4.4) is restricted to single-gate convergence cones (13% of insecure wires on Barrett). Multi-cell cones cannot be verified through our pipeline due to a limitation in Coco-Alma's per-secret SAT checking mode with multi-bit share groups. Expanding cross-validation to larger cones, through alternative SAT-based oracles, random testing, or targeted simulation, would strengthen the empirical grounding of convergence-point classifications.

**L7: MC-FM yields zero additional refinement.** Multi-cycle fresh masking analysis (MC-FM), applying FM bijection queries after MC-D1 convergence, produces zero additional wire promotions across all 7 tested modules (24 to 143K cells, spanning all module categories). This negative result has a structural explanation: FM bijection queries test whether a single randomness bit acts as a one-time pad on a wire's output, but once labels have propagated through DFF boundaries, the relevant randomness bits are no longer in the combinational fan-in of the query, they were consumed in earlier pipeline stages. The bijection property degrades across sequential boundaries because the randomness-to-output mapping spans multiple clock cycles, which combinational FM queries cannot model. This is not a dead-end but an insight: effective multi-cycle refinement requires tracking randomness provenance across register stages, not merely re-applying single-cycle bijection checks to multi-cycle labels. SADC partially addresses this limitation: by reparametrizing per-wire rather than per-randomness-bit, the value-independence check is well-defined whether or not the randomness was consumed upstream, as long as the wire's combinational cone exposes the relevant share bits to the solver.

**L9: Combinational analysis and mask-provenance assumptions.** SADC has two scope assumptions that must be stated explicitly. First, the analysis evaluates wires in their combinational cones independent of control-path reachability. This is conservative: an unreachable cone correctly remains in the upper bound, so the candidate-insecure verdict is sound regardless of whether the wire is ever exercised by the surrounding control logic. Promoting wires *out* of the upper bound by exploiting reachability would require explicit control-flow constraint injection, which is future work.

Second, the arithmetic consumer of masking (Barrett) is verified under the abstract Assumption (A2) of Theorem 3.9.1: that the arithmetic mask $s_1$ is delivered uniformly on $\mathbb{Z}_q$ and independently of the secret $x$. Our SECURE verdicts certify the consumer's arithmetic logic under this assumption; end-to-end first-order security against a power-analysis adversary additionally requires that the upstream producer of arithmetic shares (the RNG pathway and any share-generation module) respect (A2) in physical implementation. Lemma T5 of the formal proof suite (Section 3.9.5) quantifies how close a raw $k$-bit RNG comes to (A2) when it is reduced modulo q to produce an arithmetic share: for $k = 12$ and $q = 3{,}329$ the maximum bias ratio is 2 (residues 0–766 appear with probability $2/4096$, residues $767$–$3328$ with probability $1/4096$). Importantly, this hypothetical 12-bit-RNG-reduced-modulo-$q$ scenario is *distinct from* the `rnd_12bit` randomness port of the Adams Bridge `masked_barrett_reduction` module discussed in §4.5: that port is used as a raw Boolean-XOR mask (uniformity over $2^{12}$, which is the correct requirement for XOR masking and needs no rejection sampling). T5 is the general bound for deployers who might generate arithmetic shares by reducing an RNG output modulo $q$, regardless of whether Adams Bridge itself does so. A deployer whose share-generation pathway *does* apply such a raw reduction without rejection sampling has invalidated (A2) by at most the T5 bias ratio, and our SECURE verdicts must be re-interpreted accordingly; the paper does not claim end-to-end security in that regime. Strengthening SADC itself to carry explicit mask-bias bounds through to the verdicts is left to future work.

The analysis also follows the standard verification-tool convention that the *consumer* of masking is the verification target, and the *producer* of masking (the upstream RNG and share-generation pathway) delivers shares according to a labeling contract. SADC assumes that the inputs declared as randomness in the input port map are uniform and pairwise independent within their declared domain: $\{0,1\}^k$ for Boolean masks, $\mathbb{Z}_q$ for arithmetic masks. This is the same scope split adopted by SILVER, Coco-Alma, maskVerif, and Prover; none of those tools re-verifies the upstream RNG of the design under test.

For Adams Bridge ML-KEM Barrett we verified, by direct inspection of the synthesised SystemVerilog port list of `masked_barrett_reduction`, that the six labeled randomness channels (`rnd_12bit`, `rnd_14bit`, `rnd_24bit`, `rnd_for_Boolean0`, `rnd_for_Boolean1`, `rnd_1bit`) match the hardware module declaration one-to-one, there is no unmodeled randomness port. The `rnd_12bit`, `rnd_14bit`, and `rnd_for_Boolean*` channels appear in the design as raw $2^k$-bit XOR (Boolean) masks, where uniformity over $2^k$ is exactly the correct requirement and no rejection sampling is needed. The 24-bit arithmetic input shares to Barrett (`x[0+:24]` and `x[24+:24]`) are *not* generated inside the module; they are arithmetic shares produced by upstream Adams Bridge logic and consumed by Barrett. The SADC arithmetic value-independence query enforces the constraint $S_1 < q$ symbolically, so the SECURE verdicts are mathematically sound for callers that deliver shares uniformly on $\mathbb{Z}_q$ (via rejection sampling or equivalent).

Under raw $2^k$-bit unreduced randomness without rejection sampling, a $q = 3{,}329$ share sampled from a 12-bit RNG has a maximum bias ratio of $\lceil 4096/3329 \rceil = 2$ between over- and under- represented residues (residues $0$ through $766$ are sampled with probability $2/4096$ each, residues $767$ through $3328$ with probability $1/4096$). Whether the upstream Adams Bridge code base applies rejection sampling at the share-generation interface is a property of the producer module, not of Barrett, and is outside the verification scope of this paper. Strengthening SADC itself to model mask non-uniformity directly, so that the labeling contract can carry explicit bias bounds and the verdicts can be qualified with those bounds, is straightforward in principle but increases query size and is left as future work.

**L10: Direct Coco-Alma comparison on the full Barrett module.** A stable per-location Coco-Alma run on the full 5,543-cell ML-KEM Barrett extraction is left to future work. Existing Coco-Alma artifacts target a simpler extraction with a single randomness channel; building a matching Verilator testbench and VCD trace for the 6-channel 5,543-cell module is engineering-bounded but not methodologically blocked. The §4.6 butterfly subsection provides a small-scale Coco-Alma comparison on a 333-cell extraction where the two tools agree on definition-shared cells; extending this to Barrett would provide a cross-tool calibration of the 198 / 165 SADC split.

**Future work summary.** The most promising directions are: (1) distributional arithmetic SADC closing the L1 value-independence gap on the 165 Barrett residual; (2) glitch-extended SADC tracking transient values during combinational settling; (3) higher-order extension to $d \geq 2$ probing; (4) compositional verification exploiting module hierarchy and PINI/SNI guarantees; (5) the direct Coco-Alma calibration on full Barrett deferred under L10; (6) integration with TVLA-based empirical validation to correlate structural and SADC flags with measured leakage; (7) lifting T1 of the formal proof suite (`scripts/proofs/`) from its current small-domain finite expansion ($q = 5$, the entire wire-function space at that modulus) to a fully-universal mechanization quantifying simultaneously over all primes $q$ and the entire wire-function space, this is the only methodological gap remaining in the Section 3.9 proof suite, and would bring Section 3.9 backbone proof to a status comparable to the universal T1–T15 of [42].

## 7. Conclusion

We have shown that sound first-order masking verification can scale from the few-thousand-cell gadget regime of exact tools (SILVER, Prover, Coco-Alma) to production-scale arithmetic modules in real post-quantum hardware. The four-stage hierarchy, D0/D1 structural dependency analysis, fresh-mask refinement (FM), Boolean SADC, and arithmetic SADC over a symbolic modular-subtraction reparametrization, resolves 198 of 363 structurally-flagged wires (54.5%) on the 5,543-cell ML-KEM Barrett reduction module of Adams Bridge as machine-verified secure under value-independence, narrows the residual to 165 candidate first-order insecure wires as a sound upper bound, and produces every verdict in approximately three minutes on a single core with full Z3 + CVC5 dual-solver agreement on every query (363 of 363 wires, 0 disagreements). On the DOM AND reference gadget the same pipeline eliminates 100% of structural false positives in under 0.1 seconds. The result is the kind of fast, explainable, designer-actionable feedback loop that PQC hardware teams need when iterating partial-masking strategies on production NTT accelerators: hundreds of structural flags become 198 secure certificates plus 165 actionable candidates, with no indeterminate residual and no deferral to external exact tools.

The pipeline complements concurrent work on partial-NTT-masking security margins [42]. Where that paper quantifies which INTT layers must be masked to preserve a target security margin against soft-analytical side-channel attacks, this paper provides the SMT-based machinery to verify that the chosen masking is correct at the gate level. Together they form an emerging design loop for partial-masking PQC hardware: choose a masked layer set with a quantified margin, verify the chosen gates structurally and distributionally, iterate.

Three independent methods, structural verification (this work), SAT-based exact analysis (Coco-Alma), and physical measurement [9], converge on the same architectural pattern in Adams Bridge, illustrating that the SADC pipeline is complementary to, not a replacement for, exact algebraic verification and empirical measurement. As FIPS 140-3 certification increasingly encompasses post-quantum implementations, scalable pre-silicon tools that offer sound coverage and machine- checkable secure certificates can support the verification chain.

Regarding the limitations, arithmetic SADC currently uses value-independence as a *sufficient* condition for distributional security (Section 3.9.6, and 6, L1); the 165 candidate-insecure wires on Barrett are therefore a sound over-approximation that a full distributional check (e.g., model counting under uniform-mask assumptions, or threshold- implementation rebalancing analysis) may further reduce. The combinational cone analysis evaluates wires independent of control- path reachability (Section 6, L9), which is conservative: any unreachable cone correctly remains inside the upper bound but cannot be promoted out of it without explicit reachability constraints. The Barrett results assume the six labeled randomness channels are uniform and pairwise independent within their declared domain; under raw $2^k$-bit unreduced randomness without rejection sampling, the verified-secure guarantees hold only under a negligible-bias assumption on the residual mod-$q$ distribution. A direct stable per-location Coco-Alma comparison on the full arithmetic Barrett extraction is left to future work (Section 6, L10): existing Coco-Alma artifacts target a simpler single-randomness- channel extraction, and a matching Verilator testbench for the 6-channel 5,543-cell module is engineering-bounded rather than methodologically blocked.

Three directions could extend this paper: (i) closing the value-independence gap via distributional model counting on the 165 Barrett residual; (ii) higher-order extension of arithmetic SADC for $d \geq 2$ probing adversaries; (iii) a stable per- location Coco-Alma baseline on the full 6-channel Barrett extraction for cross-tool calibration of the 198 / 165 split. The tool, test suite, evidence files, and `repro_manifest.json` listing all version pins, random seeds, and resource limits used in this evaluation will be made available for TCHES artifact evaluation. A pure-Python Z3-free Boolean evaluator, used in §4.6 for ground-truth re-checking of small SADC verdicts, is bundled with the release.